\documentclass[10pt,notitlepage,nofootinbib,preprintnumbers,showpacs]{revtex4-2}
\usepackage[english]{babel}
\pdfoutput=1

\usepackage{hyperref}
\usepackage{color,graphicx,epsfig} 
\usepackage{ifpdf}
\usepackage{tabularx}
\usepackage{amsmath} 
\usepackage{amssymb}
\usepackage{bm}
\usepackage[utf8]{inputenc}

\usepackage{array}

\usepackage{graphics}
\usepackage{longtable}
\usepackage{slashed}
\usepackage{braket}
\usepackage{lineno}

\makeatletter

\usepackage{textcomp}
\usepackage[usenames,dvipsnames]{xcolor}
\usepackage{relsize}
\usepackage{bbm}
\usepackage{tikz}
\usepackage{tikz-feynman}

\graphicspath{{plots/}}

\usetikzlibrary{calc,patterns,angles,quotes,arrows}

\usepackage{pifont}
\usepackage{multirow}

\allowdisplaybreaks
\newcommand{\nn}{\nonumber}
\newcommand{\e}[1]{\mathrm{\,#1}}
\newcommand{\re}[0]{\mrm{Re}}
\newcommand{\im}[0]{\mrm{Im}}
\newcommand{\E}[1]{\times 10^{#1}}       

\newcommand{\nc}{\newcommand}
\nc{\Xx}[1]{X^{(#1)}}
\nc{\cx}[1]{c^{(#1)}}
\nc{\mrm}{\mathrm}
\nc{\mc}{\mathcal}
\nc{\non}{\nonumber}
\nc{\hc}{\hbox {H.c.}}
\nc{\noi}{\noindent}
\nc{\barx}{\bar{x}}
\nc{\pbarn}{\;\hbox {pb}}
\nc{\fbarn}{\;\hbox {fb}}

\nc{\hsp}{\hspace{0.5cm}}
\nc{\lsp}{\hspace{1cm}}
\nc{\Lsp}{\hspace{2cm}}
\nc{\LLsp}{\lsp\lsp}
\nc{\lra}{\longrightarrow}
\nc{\p}{\prime}
\nc{\sgn}{\text{sgn}}
\nc{\ph}{\varphi}
\nc{\op}{{\cal O}}
\nc{\eq}{\text{Eq.~}}

\nc{\beq}{\begin{equation}}  \nc{\eeq}{\end{equation}}
\nc{\bea}{\begin{eqnarray}}  \nc{\eea}{\end{eqnarray}}
\nc{\baa}{\begin{array}}     \nc{\eaa}{\end{array}}
\nc{\bit}{\begin{itemize}}   \nc{\eit}{\end{itemize}}
\nc{\ben}{\begin{enumerate}} \nc{\een}{\end{enumerate}}
\nc{\bce}{\begin{center}}    \nc{\ece}{\end{center}}
\nc{\bpm}{\begin{pmatrix}}   \nc{\epm}{\end{pmatrix}}
\nc{\br}{\mc{B}}

\makeatother

\definecolor{niceblue}{rgb}{0.15,0.15,0.6}
\definecolor{nicegreen}{rgb}{0.1,0.5,0.1}
\definecolor{nicered}{rgb}{0.7,0.1,0.1}
\definecolor{violet}{rgb}{0.7,0.3,0.3}

\definecolor{Red}{rgb}{1.,0.,0.}

\definecolor{Green}{rgb}{0.2,.7,0.2}

\nc{\jfk}[1]{\textcolor{nicered}{{\bf JFK:} #1}}
\nc{\nk}[1]{\textcolor{violet}{{\bf NK:} #1}}

\hypersetup{colorlinks,citecolor= nicegreen,linkcolor= niceblue}
\hypersetup{colorlinks=true}
\begin{document}

\author{Svjetlana Fajfer} \email[Electronic
address:]{svjetlana.fajfer@ijs.si} 
\affiliation{Department of Physics,
  University of Ljubljana, Jadranska 19, 1000 Ljubljana, Slovenia}
\affiliation{J. Stefan Institute, Jamova 39, P. O. Box 3000, 1001
  Ljubljana, Slovenia}

\author{Jernej F. Kamenik} 
\email[Electronic address:]{jernej.kamenik@ijs.si}
\affiliation{Department of Physics,
  University of Ljubljana, Jadranska 19, 1000 Ljubljana, Slovenia}
\affiliation{J. Stefan Institute, Jamova 39, P. O. Box 3000, 1001 Ljubljana, Slovenia}

\author{Arman Korajac} 
\email[Electronic address:]{arman.korajac@ijs.si}
\affiliation{Department of Physics,
  University of Ljubljana, Jadranska 19, 1000 Ljubljana, Slovenia}
\affiliation{J. Stefan Institute, Jamova 39, P. O. Box 3000, 1001 Ljubljana, Slovenia}

\author{Nejc Ko\v snik} 
\email[Electronic address:]{nejc.kosnik@ijs.si}
\affiliation{Department of Physics,
  University of Ljubljana, Jadranska 19, 1000 Ljubljana, Slovenia}
\affiliation{J. Stefan Institute, Jamova 39, P. O. Box 3000, 1001 Ljubljana, Slovenia}

%
%
\title{Correlating New Physics Effects in Semileptonic $\Delta C=1$ and $\Delta S=1$ Processes}
%
%

\date{\today}

\begin{abstract}
  We present constraints on the left-handed dimension-6 interactions that contribute to semileptonic and leptonic decays of $K$, $D$, pions and to nuclear beta decay. We employ the flavour covariant description of the effective couplings, identify universal CP phases of New Physics and derive constraints from decay rates and CP-odd quantities. As a result, we can predict the maximal effects of such flavoured NP in $D$ decays from stringent $K$ decay constraints and vice-versa.
\end{abstract}


\maketitle

\unitlength = 1mm

%
\section{Introduction}
\label{sec:intro}
%

The Standard Model (SM) has a unique way of incorporating CP violation
(CPV) and suppressing flavor changing neutral currents (FCNCs) in the
quark sector. In particular, the lightness of the first two generations and the suppressed mixing with the third generation severely suppress FCNC transitions involving only the first two generation quarks. 
This is manifest in particular in processes which are characterized by so called hard GIM, such as those involving CPV. 
In fact, no significant deviations from the SM predictions related to
strangeness ($S$) and charm ($C$) flavor violation have been observed in
experiments to date, placing stringent limits on possible beyond the SM (BSM) effects in these sectors.

On the other hand, observed hints of deviations from SM predictions in semileptonic $\Delta B=1$
processes (both in $b \to s \mu \mu$ FCNCs and in particular
in $b \to c \tau \nu$ charged current (CC) mediated semileptonic $b$-hadron decays), still await experimental clarification, and have triggered many
studies throughout the last decade (see
e.g. Refs.~\cite{Bifani:2018zmi,Artuso:2022ijh}).
Intriguingly, the most straightforward and successful BSM proposals
addressing the $\Delta B=1$ FCNC and CC observables introduce new interactions of left-handed quarks (and leptons)~\cite{Buttazzo:2017ixm, Cornella:2019hct, Barbieri:2022ikw, Angelescu:2021lln, Dorsner:2017ufx, Becirevic:2018afm, Becirevic:2022tsj, Crivellin:2017zlb}, which imply novel flavor breaking sources (besides SM Yukawas) of the $U(3)_Q$ flavour symmetry, respected by the SM gauge interactions. This already motivated a reconsideration of BSM effects in (rare) (semi)leptonic decays of kaons~\cite{Aebischer:2022vky, Buras:2022irq}, $D$-mesons~\cite{Fajfer:2015mia,DeBoer:2018pdx,Bause:2020xzj}, and also top quarks~\cite{Kamenik:2018nxv}.

Experimentally, there has been recent progress in the search for rare $\Delta C=1$
leptonic $D^0\to \mu^+ \mu^-$~\cite{LHCb:2022uzt} decay as well as the
analysis of non-resonant regions of the differential rate for
$D^+ \to \pi^{+} \mu^+ \mu^-$~\cite{LHCb:2020car} by the LHCb
collaboration. BESIII collaboration has also recently reported results
from a first dedicated search for $D^0 \to \pi^0 \nu \bar \nu$
decay~\cite{BESIII:2021slf}.  Similarly, new results have been
recently reported on semileptonic $\Delta S=1$ transitions in both
charged~\cite{NA62:2021zjw, NA62:2022qes} and
neutral~\cite{KOTO:2020prk} kaon decays by the NA62 and KOTO
collaborations, respectively. Further significant improvements in
these measurements and searches are expected from these and the next
generation of flavor
experiments~\cite{EuropeanStrategyforParticlePhysicsPreparatoryGroup:2019qin}.

Motivated by these developments, we investigate the interplay of
possible NP effects in semileptonic CC and FCNC transitions involving
purely left-handed first- and second-generation quarks. In particular,
it has been shown previously~\cite{Gedalia:2012pi} that the peculiar
structure of $U(3)_Q$ breaking in the SM implies that possible BSM
sources of CPV in this sector affect rare charm and kaon decays in a
universal way (see also Ref.~\cite{Blum:2009sk}). We demonstrate how
and when existing bounds on CPV in rare semileptonic $K$ meson decays
severely constrain the possible size of the corresponding effects in
charm decays, and vice-versa. Employing the covariant parametrization
of flavor conversion developed in Refs.~\cite{Blum:2009sk, Gedalia:2010zs, Gedalia:2010mf} we constrain the unique new CPV
parameter whose effect cannot be tuned by adjusting the alignment
angle of BSM flavor breaking to down-quark or up-quark basis. Furthermore, we derive robust
model-independent bounds on BSM affecting either charm or kaon
semileptonic decays, and discuss the interplay between CC and FCNC
transitions. Finally, we study the increasingly important constraints posed by the
experimental studies of high-$p_T$ semileptonic processes at the LHC
$pp \to \ell \nu (\ell^+\ell^-)$~\cite{Fuentes-Martin:2020lea,Allwicher:2022gkm,Greljo:2022jac}.

The remainder of the paper is structured as follows: in Sec.~\ref{sec:framework} we review the basic elements of the SM effective theory (SMEFT) of flavour conversion including CPV within the first two generations of left-handed quarks. We apply this framework to (rare) semileptonic $K$ and $D$ meson decays in Sec.~\ref{sec:KD}. Secs.~\ref{sec:sdnunucull} and~\ref{sec:sdllcununu} contain the detailed discussion of the relevant observables connecting and constraining the semileptonic $\Delta C=1$ and $\Delta S =1$ FCNC processes $s\to d \nu \bar \nu$ and $c\to u \ell^{+} \ell^{-}$, and $s\to d \ell^{+} \ell^{-}$ and $c\to u  \nu \bar \nu$, respectively. We explain the interplay between the two sectors in high-$p_T$ collider experiments in Sec.~\ref{sec:HighPTlimits} and discuss the additional correlations introduced by the inclusion of CC processes in Sec.~\ref{sec:CC}. Sec.~\ref{sec:results} contains our main results and projections, while we present our conclusions and prospects for future experiments in Sec.~\ref{sec:conclusions}.

%
\section{Framework}
\label{sec:framework}
%
We are interested in BSM effects in semileptonic transitions involving
exclusively left-handed quarks of first two generations. Working within
the SM effective field theory (SMEFT)~\cite{Grzadkowski:2010es} valid below a heavy new physics (NP) threshold scale $\Lambda$, we
thus supplement the SM Lagrangian by local
semileptonic effective operators with left-chiral
quarks\footnote{Additional SMEFT operators modifying $W$ and $Z$
  couplings can also contribute, however they are constrained by precision
  measurements of on-shell massive weak vector bosons at
  LEP~\cite{ALEPH:2005ab,0708.1311,ParticleDataGroup:2022pth}.}
  \begin{equation}\label{eq:lsmeft}
  	\mathcal{L}_{\mathrm{SMEFT}} \supset
  	\frac{X^{(3,\ell)}_{ij}}{\Lambda^2} (\bar Q_i \gamma_\mu
        \sigma^a Q_j) (\bar L_\ell \gamma^\mu \sigma_a
        L_\ell)+\frac{X^{(1,\ell)}_{ij}}{\Lambda^2} (\bar Q_i
        \gamma_\mu Q_j) (\bar L_\ell \gamma^\mu L_\ell)\,.
  \end{equation}
  Here $Q_i$ is the $i$-th generation left-handed quark doublet, which
  we write in the down-quark mass basis as
  $Q_i = ( u'_{Li}, d_{Li})^T$. The up-quark fields in this basis are
  related to their mass eigenstates via the CKM matrix $V$ as
  $u_i' = V^*_{ji} u_{j}$. For leptons we choose the charged lepton
  mass basis: $L_i = (U^*_{ji }\nu_{Lj}, \ell_{Li})^T$, where $U$ is
  the PMNS matrix. Pauli matrices $\sigma^a$, $a=1,2,3$, act in the $\mrm{SU}(2)_L$
  space. We assume in Eq.~\eqref{eq:lsmeft} that lepton flavour is
  conserved, whereas the BSM quark flavour conversion is parametrized
  by Hermitian matrices $X^{(1,\ell)}, X^{(3,\ell)}$.
  The resulting Lagrangian containing FCNCs reads
  \begin{equation}
     \label{eq:FCNCs}
    \begin{split}
      \mc{L}_\mrm{FCNC} &= \frac{1}{\Lambda^2} X_{ij}^{(+)} \left[ (\bar u_i' \gamma^\mu P_L u_j') (\bar \nu \gamma_\mu P_L \nu) + (\bar d_i \gamma^\mu P_L d_j) (\bar \ell \gamma_\mu P_L \ell) \right]  \\
      &+  \frac{1}{\Lambda^2} X_{ij}^{(-)} \left[ (\bar u_i' \gamma^\mu P_L u_j') (\bar \ell \gamma_\mu P_L \ell) + (\bar d_i \gamma^\mu P_L d_j) (\bar \nu \gamma_\mu P_L \nu) \right] \,,
    \end{split}
  \end{equation}
  where $P_{R,L} = (1\pm\gamma_5)/2$. Above, we have introduced the
  matrices $X^{(\pm)} = X^{(1)} \pm X^{(3)}$ and suppressed explicit
  lepton flavour index for clarity.  On the other hand, the charged
  currents stemming from Eq.~\eqref{eq:lsmeft} are only due to the
  $X^{(3)}$
  \begin{equation}
    \label{eq:CCs}
    \begin{split}
      \mc{L}_\mrm{CC} &= \frac{1}{\Lambda^2}\,2 X_{ij}^{(3)} (\bar u_i' \gamma^\mu P_L d_j) (\bar \ell \gamma_\mu P_L \nu) + \mrm{h.c.} \,.
    \end{split}
  \end{equation}

  Next we focus exclusively on the first two generations and use the
  fact that any two-dimensional hermitian matrix can be decomposed in
  terms of the identity and Pauli matrices. Note that in isolating the
  first two generations in the following we are neglecting possible
  additional BSM effects due to mixing with the third quark
  generation. However, the resulting modifications of our results are
  in general severely suppressed due the hierarchical structure of the
  SM quark Yukawas. See Ref.~\cite{Gedalia:2012pi} for in depth
  discussion on this point. We can write
  \begin{equation}
    \label{eq:decomposition}
    X^{(\pm)}_{ij} = \lambda^{(\pm)} \delta_{ij} +  \cx{\pm}_a (\sigma^a)_{ij}\,,
  \end{equation}
 where $\lambda$ and $c_a$ are real. It is only the traceless part ($c_a$) that plays a role in FCNC processes. In contrast, $\lambda$'s contribute to flavour-diagonal neutral currents as well as to charged current processes via $\Xx{3}$:
  \begin{equation}
    \label{eq:C3}
    2 \Xx{3}_{ij} = (\lambda^{(+)} - \lambda^{(-)}) \delta_{ij} +  (\cx{+}_a - \cx{-}_a) (\sigma^a)_{ij} \,.
  \end{equation}
  Notice that a unique parameter, $c^{(\pm)}_2$, encodes CP violation, while the remaining three couplings are real. The
  traceless part of the coupling matrix offers an intuitive  geometrical interpretation~\cite{Gedalia:2010mf} since it spans a
  3-dimensional space. Each traceless hermitian matrix $A$ is equivalent to a  real 3-dimensional vector $\bm{a}$ via the mapping
  $A = \bm{a} \cdot \bm{\sigma}$. Scalar and cross product  between vectors $\bm{a}$, $\bm{b}$ (corresponding to matrices
  $A = \bm{a}\cdot \bm{\sigma}, B = \bm{b}\cdot \bm{\sigma}$) are  defined via matrix operations as
  \begin{align}
    \label{eq:euclidean}
    \bm{a} \cdot \bm{b} \equiv \frac{1}{2}\mrm{Tr} [A B],\qquad
    \bm{a} \times \bm{b} \equiv \frac{-i}{2} [A, B],
  \end{align}
  and allow for interpretation in terms of lengths, angles and
  volumes.  
  
\begin{figure}[!h]
  \centering
  		\begin{tikzpicture}[
			scale=2,
			axis/.style={thin, ->, >=stealth'},
			important line/.style={very thick, ->, >=stealth'},
			every node/.style={color=black, opacity=1.}
			]
			\coordinate (origin) at (0.,0.);
			\coordinate (poly1) at (1.73, 1.0);
			\coordinate (poly2) at (3.73, 1.0);
			\coordinate (poly3) at (2., 0.0); 
			\coordinate (poly4) at (1.4142, 1.4142);
			\coordinate (poly5) at (3.1442, 2.4142); 
			\coordinate (poly6) at (1.73, 1.0); 
			\coordinate (origo) at (0.25,0.);
			\coordinate (pivot1) at (10.,5.8);
			\coordinate (pivot2) at (10.,0);
                        \coordinate (pivot3) at (10.,10);
			\draw[black, ultra thin, fill=orange, fill opacity=0.4] (0.,0.) -- (1.73, 1.0) -- (3.73, 1.0) -- (2., 0.0) -- cycle;
			\draw[black, ultra thin, fill=blue, fill opacity=0.4] (0.,0.) --(1.4142, 1.4142) -- (3.1442, 2.4142) -- (1.73, 1.0) -- cycle;
                        \draw[axis] (-0.1,0)  -- (2.5,0) node(xline)[right]
                        {$- \sigma_3$};
                        \draw[axis] (0,-0.1) -- (0,2.5) node(yline)[above] {$\sigma_1$};
			\draw[important line] (0,0.) -- (1.4142, 1.4142) node(zline)[above] {$\hat{\mathcal{A}}_u$};
			\draw[important line] (0,0.) -- (2., 0.0) node(zline)[below] {$\hat{\mathcal{A}}_d$};
			\draw[important line] (0,0.) -- (1.73, 1.0) node(zline)[above right] {};
			\pic [draw, -, angle radius=18mm, angle eccentricity=1.2, "$\theta_d$"] {angle = pivot2--origin--pivot1};
                        \pic [draw, -, angle radius=25mm, angle eccentricity=1.2, "$2\theta_c$"] {angle = pivot2--origin--pivot3};
			\node (none) at (1.56,1.2) {$|z_{\Delta C = 1}|$};
			\node (none) at (2.1,0.52) {$|z_{\Delta S = 1}|$};
			\node (none) at (2.13,1.14) {$X$};
                        \node (none) at (1.6,0.78) {$c_R$};
                        \draw [thick] (-0.02,-0.02) circle [radius=0.05];
                        \draw [fill] (-0.02,-0.02) circle [radius=0.01];
                        \node (none) at (0.2,-0.2) {$c_I \sigma_2 \sim \hat{\mathcal{A}_d} \times \hat{\mathcal{A}_u}$};
                      \end{tikzpicture}
  \caption{Schematic of possible NP contributions to $ \Delta S=1$ and $\Delta C =1$ FCNC
    semileptonic processes in the two-generation limit of SM. CP conserving magnitudes of NP contributions ($|z_{\Delta C=1}|,\,|z_{\Delta S = 1}|$) depend on the alignment angle $\theta_d$. CPV NP contributions ($c_I$) are independent of $\theta_d$. See text for details.}
  \label{fig:alignment}
\end{figure}
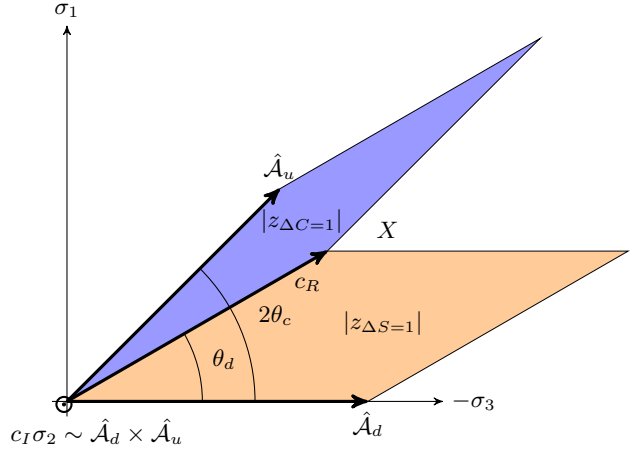

  Our analysis is based on the SM flavour group for the
  first two quark generations
  $\mathcal F=U(2)_{Q} \times U(2)_{U} \times U(2)_{D}$ where $Q$, $U$
  and $D$ stand for quarks doublets, up-type singlets and down-type
  singlets, respectively \cite{Blum:2009sk,Gedalia:2010mf}.  The group
  $\mathcal F$ is broken within the SM only by the Yukawa
  interactions.
To better understand the resulting pattern of flavour and CP violation in and beyond the SM, we promote  $Y_{u}$
and $Y_{d}$ to spurions that transform under $\mathcal{F}$  as
$(2,2,1)$ and $(2,1,2)$, respectively. 
In order to construct $\mc{F}$-invariant terms
we furthermore define the spurions $Y_u Y_u^{\dagger}$ and $Y_d
Y_d^{\dagger}$. They belong to the $(3 \oplus 1, 1, 1)$ representation
of  $\mathcal F$. Since the traces of these matrices do not affect
flavour-changing processes, it is useful to remove them and work with traceless parts:
\begin{equation}
  \mathcal{A}_{u,d} = (Y_{u,d}^{\phantom{\dagger}} Y_{u,d}^{\dagger})_\slashed{\mrm{tr}} \,,\quad \textrm{where } M_\slashed{\mrm{tr}} \equiv M - \frac{1}{2} \mathbb{I}\,\mrm{tr} M\,.
\end{equation}
Both $\mc{A}_{u,d}$ belong to the adjoint representation of $U(2)_{Q}$.  Without loss of
generality, we can choose to work in a basis with diagonal down-quark Yukawa couplings, $Y_d = \mathrm{diag} (y_d, y_s)$, whereas $Y_u = V_\mathrm{CKM} \mathrm{diag} (y_u, y_c)$. There is no CP violation in the SM with two generations, leading to $V_\mrm{CKM}$ being just a rotation matrix with the Cabibbo angle $\theta_{c}$.  Inserting $Y_u$ and $Y_d$, we get:
\begin{align}
\mathcal{A}_{d} &= \frac{y_d^2 - y_s^2}{2}  \sigma_3 ,\\
\mathcal{A}_{u} &= V \begin{pmatrix}
	y_u & 0 \\
	0 & y_c 
\end{pmatrix}^2 V^T
= \frac{y_c^2 - y_u^2}{2}\Big(-\cos(2\theta_c) \sigma_3 + \sin(2 \theta_c) \sigma_1\Big). 
\end{align}
For later convenience, we introduce normalised vectors
  \begin{eqnarray}
\hat{\mathcal{A}}_{d} &=& - \sigma_3 \label{a_down}\,, \\
\hat{\mathcal{A}}_u & = & - \cos (2\theta_c) \sigma_3 + \sin (2 \theta_c) \sigma_1\,,
\end{eqnarray}
which are shown in Fig.~\ref{fig:alignment}. These two basis vectors present two special directions in the coupling
space of $X$, since alignment of $X$ along one of them,
$X \propto \mc{A}_{u,d}$, implies no FCNCs in up-type or down-type
quarks, respectively. However, BSM couplings can also span the orthogonal direction
 along $\sigma_2$.
The most general form of $X$ that  includes this CPV direction can thus be written as $X_\slashed{\mrm{tr}} = \alpha \hat{\mc{A}}_u + \beta \hat{\mc{A}}_d + i \gamma [\hat{\mc{A}}_u, \hat{\mc{A}}_d]$.
In the remainder of this paper, we shall use cylindrical coordinates
$c_R, c_I,$ and $\theta_d$, which are related to the cartesian ones as
$c_1 = c_R \sin\theta_d$, $c_3 = -c_R \cos\theta_d$ and $c_2 =
c_I$. The most general form of $X$ then reads
\begin{align}
  \label{eq:cylindrical}
  X &= \lambda \mathbb{I} +c_R \sin\theta_d  \,\sigma_1  + c_I\, \sigma_2 - c_R \cos\theta_d\, \sigma_3\,.
\end{align}
We conclude this section by commenting on the approximations taken in the two generation limit of the SM. Within the full three-generation SM (as well as in minimally flavor violating (MFV) NP scenarios~\cite{DAmbrosio:2002vsn}) CPV effects in flavor changing processes among the first two generations are not strictly vanishing, but are nonetheless severely suppressed by small CKM mixing with the third generation. Within our framework, such suppressed effects would lead to $\mathcal A_{u,d}$ acquiring small components in the $\sigma_2$ direction in flavour space. The NP flavor alignment limit with either up- or down-type quark mass basis would then imply alignment of NP and SM CPV phases as well. In our analysis we are neglecting these subleading, suppressed CPV effects of NP and are thus effectively probing (CPV) NP effects beyond MFV.

%
\section{Aligning BSM flavour structures with $\Delta S=1$ and $\Delta C=1$ constraints}
\label{sec:KD}
%
We first focus on the allowed size of the CP-even
$c_{R}$ and CP-odd $c_I$ couplings depending on the alignment angle
$\theta_d$. At low energies, the $\Xx{\pm}$ matrices map onto parameters of the
effective Lagrangian
\begin{align}
  \label{eq:Llow}
  \mc{L}_\mrm{eff}^{(\pm)} &= \frac{z_{\Delta S=1}^{(-)}}{\Lambda^2} (\bar d_L \gamma^\mu s_L)(\bar \nu_L \gamma_\mu \nu_L)
                             + \frac{z_{\Delta C=1}^{(-)}}{\Lambda^2} (\bar u_L \gamma^\mu c_L)(\bar \ell_L \gamma_\mu \ell_L)  +\\
   &+ \frac{z_{\Delta S=1}^{(+)}}{\Lambda^2} (\bar d_L \gamma^\mu s_L)(\bar \ell_L \gamma_\mu \ell_L)
                             + \frac{z_{\Delta C=1}^{(+)}}{\Lambda^2} (\bar u_L \gamma^\mu c_L)(\bar \nu_L \gamma_\mu \nu_L)  \nn
+  \mrm{h.c.}\,,
\end{align}
written in the fermion mass basis. The magnitudes $|  z^{(\pm)}_{\Delta S=1} |,\,|  z^{(\pm)}_{\Delta C=1} |$ can be geometrically expressed in a
manifestly basis-independent form:
\begin{align}
  \label{eq:zAbs}
  |z_{\Delta S=1}^{(\pm)}| &= |\Xx{\pm} \times \hat{\mathcal{A}}_{d}| = \sqrt{\left(c_R^{(\pm)} \sin \theta_d^{(\pm)}\right)^2 + (c_I^{(\pm)})^2}\,,\\
  |z_{\Delta C=1}^{(\pm)}| &= |\Xx{\pm} \times \hat{\mathcal{A}}_{u}|  = \sqrt{\left(c_R^{(\pm)} \sin (2\theta_c-\theta_d^{(\pm)})\right)^2 +(c_I^{(\pm)})^2}\,.
\end{align}
For $|z_{\Delta S=1}^{(\pm)}|$, the above form can be understood in the down-quark basis
where $\hat{\mc{A}}_d$ is proportional to $\sigma_3$. Then, $z_{\Delta S=1}^{(\pm)}$
corresponds to $X_{12}^{(\pm)}$, whose size is given by $\sqrt{(c_R \sin
  \theta_d)^2 + c_I^2}\big|_{(\pm)}$. This is exactly the length of the orthogonal component of $X^{(\pm)}$ to $\hat{\mc{A}}_{d}$ obtained by the cross-product~\cite{1003.3869}. The analogous argument holds for $|z_{\Delta C=1}^{(\pm)}|$ when we analyse it in the up-quark mass basis.
On the other hand, the CP-violating imaginary part is universal since
it is normal to the $\hat{\mc{A}}_d-\hat{\mc{A}}_u$ plane and thus insensitive to rotations in the $1-3$ plane:
\begin{align}
  \label{eq:zIm}
  \Im (z_{\Delta S=1}^{(\pm)}) &= \Im (z_{\Delta C=1}^{(\pm)}) = \Im \left(X^{(\pm)} \cdot \frac{\hat{\mc{A}}_d \times \hat{\mc{A}}_u}{|\hat{\mc{A}}_d \times \hat{\mc{A}}_u|}\right) = c_I^{(\pm)}.
\end{align}

If we rely only on the CP-even experimental upper bounds, namely if we
know only upper bounds $|z_{\Delta S=1}^\mrm{exp}|$ and $|z_{\Delta C=1}^\mrm{exp}|$,
we can minimise the effect in rare kaon decays when $\theta_d$ is small
since $X$, in this case, is aligned towards the down-quark mass
basis. Conversely, we get a minimal effect in $D$ meson decays at an angle
$\theta_d = 2 \theta_c$. At small $\theta_d$ (down-alignment) we thus
expect $\Delta C=1$ constraints to dominate, whereas for $\theta_d
\approx 2\theta_c$ (up-alignment) the $\Delta S=1$ processes become more important.
In between the two regimes lies an optimal value of angle $\theta_d^*$ at which
the constraints on $|X|$ stemming from $|z_{\Delta S=1}^\mrm{exp}|$ and
$|z_{\Delta C=1}^\mrm{exp}|$ coincide numerically, i.e. when $|z_{\Delta S=1}/z_{\Delta C=1}| = |z_{\Delta S=1}^\mrm{exp}|/|z_{\Delta C=1}^\mrm{exp}| \equiv
r_{\rm exp}$. Assuming that the alignment angle is in the range
$0 \leq \theta_d \leq 2 \theta_c$, we find
\begin{equation}
  \label{eq:thetad}
  \tan \theta_d^* = \frac{c_R^2 r_{\rm exp}^2 \sin (4 \theta_c)/\sqrt{2}- \sqrt{-2 c_I^2  (c_I^2 + c_R^2) \left(r_{\rm exp}^2-1\right)^2 +2 r_{\rm exp}^2 c_R^4 \sin^2 2\theta_c}}{\sqrt{2} \left[c_I^2
   \left(r_{\rm exp}^2-1\right)+c_R^2 r_{\rm exp}^2 \cos ^2(2 \theta_c)-c_R^2\right]}\,.
\end{equation}
In the regime of large CPV, $|c_I| \gg c_R$, there is no solution for
$\theta_d^*$, since the effect of $\theta_d$ is rendered
unimportant. The optimal alignment in the CP-conserving limit with
$c_I = 0$ reads
\begin{equation}
  \label{eq:thetadCPC}
  \tan \theta_d^*\big|_\mrm{c_I = 0} = \frac{r_{\rm exp} \sin{2 \theta_c}}{1 + r_{\rm exp}\cos{2 \theta_c}}\,.
\end{equation}

\subsection{Matching to weak effective theory}
At low energies, we rely on the weak
effective theory (WET) and use standard conventions for the
Hamiltonian governing $\Delta C=1$  transitions
\begin{equation}\label{eq:HeffUp}
  	\mathcal{H}_{\textrm{eff}}^{\Delta C=1} = - \frac{4\,G_F}{\sqrt{2}} \frac{\alpha_\mrm{em}}{4\pi}  \Big(
  	\sum_{i=9,10}
  	C_{i,\ell}^{\Delta C=1} O_{i,\ell}^{\Delta C=1}  + C^{\Delta C=1}_{L,\nu_\ell} O^{\Delta C=1}_{L, \nu_\ell}  \Big) + \text{h.c.}\,,
      \end{equation}     
  where the SMEFT NP effects are imprinted upon the following set of dimension-6 operators:
  \begin{align}
	O_{9,\ell}^{\Delta C=1} &= 
	(\bar{u} \gamma_{\mu} P_{L} c)(\bar{\ell} \gamma^\mu
                                   \ell)\,,\qquad
	O_{L, \nu_\ell}^{\Delta C=1} =
	(\bar{u} \gamma_{\mu} P_L c)(\bar{\nu}_\ell \gamma_{\mu} P_L \nu_{\ell})\,,\\                                   
	O_{10,\ell}^{\Delta C=1} &=
	(\bar{u} \gamma_{\mu} P_{L} c)( \bar{\ell} \gamma^\mu \gamma_5 \ell)\,.
\end{align}
For $\Delta S=1$ transitions, we conversely employ
\begin{equation}\label{Heffsd}
	\mathcal{H}_{\textrm{eff}}^{\Delta S=1} = - \frac{4\,G_F}{\sqrt{2}} \frac{\alpha_\mrm{em}}{4\pi}
 \Big(
\sum_{i=9,10}
C_{i,\ell}^{\Delta S=1} O_{i,\ell}^{\Delta S=1}  + C^{\Delta S=1}_{L,\ell} O^{\Delta S=1}_{L, \nu_\ell}  \Big) +  \text{h.c.}\,.
\end{equation}
The operators for the down-quark sector have the same structure as
those for the up-quark sector; they differ in a simple replacements of $u \to d$
and $c \to s$.  Here, $\ell = e, \mu$ or $\tau$. We will separate the
contribution of SM and NP to the Wilson coefficients:
\begin{equation}
	C_i = C_i^{\mathrm{SM}} + C_i^{\mathrm{NP}}\,.
\end{equation}
The left-handed SMEFT operator structure that we consider in
Eq.~\eqref{eq:FCNCs} results in the relation  $C_9^\mrm{NP} =
-C_{10}^\mrm{NP}$ for charged-lepton operators.
After matching $X^{(-)}$ SMEFT coefficients onto the WET Wilson coefficients, we find
	\begin{align}
		s \to d \nu \bar{\nu} : \qquad &C_{L, \nu}^{\Delta S=1 \mathrm{, \, NP}} = \frac{2 \pi}{ \alpha_\mrm{em}} \frac{v^2}{\Lambda^2}
		\bigg\{
		c^{(-)}_R \sin \theta^{(-)}_d - i c_I^{(-)}
                                                 \bigg\} \,,\\
                \label{eq:C9charm}
		c \to u \ell^+ \ell^- : \qquad &C_{9}^{\Delta C=1 \mathrm{, NP}}  = - C_{10}^{\Delta C=1 \mathrm{, NP}} = \frac{\pi}{ \alpha_\mrm{em}} \frac{v^2}{\Lambda^2} \bigg\{
		c_R^{(-)} \sin (\theta_d^{(-)} - 2 \theta_c)  - i c_I^{(-)}
                                                 \bigg\} \,,
        \end{align}
whereas the low-energy coefficients from $X^{(+)}$ are
\begin{align}
		s \to d \ell^+ \ell^- : \qquad &C_{9}^{\Delta S=1 \mathrm{, NP}}  = - C_{10}^{\Delta S=1 \mathrm{, NP}} = \frac{\pi}{ \alpha_\mrm{em}} \frac{v^2}{\Lambda^2}
		\bigg\{
		c_R^{(+)} \sin \theta_d^{(+)} - i c_I^{(+)}
		\bigg\} \,,\\
		c \to u \nu \bar{\nu} : \qquad	&C_{L, \nu}^{\Delta C=1 \mathrm{,  NP}} = \frac{2 \pi}{ \alpha_\mrm{em}} \frac{v^2}{\Lambda^2}
		\bigg\{
		c_R^{(+)} \sin (\theta_d^{(+)} - 2 \theta_c) - i c_I^{(+)}
		\bigg\} \,.
	\end{align}
The presented Wilson coefficients indicate how the CP conserving NP contributions to 
charm and kaon physics are related via the
Cabibbo rotation and its interplay with the alignment angle. In the
remainder of the paper we study current constraints on
$X^{(+)}$ and $X^{(-)}$, as parameterised by $c_R^{(\pm)}$,
$\theta_d^{(\pm)}$ and $c_I^{(\pm)}$.
The presence of $c_I^{(\pm)}$ without any $\theta_d$ dependence
implies the flavor universal character of the CPV parameters.

In our numerical studies we set the scale to $\Lambda =
1\e{TeV}$, thus all the presented bounds on $c_{R,I}$ should be understood as
bounds on $(\mrm{TeV}/\Lambda)^2 c_{R,I}(\Lambda)$.\footnote{The
renormalization group running effects of the left-handed semileptonic operators are negligible~\cite{Jenkins:2013wua}.}

%
\section{$s \to d \nu \bar \nu$ and $c \to u \ell^+ \ell^-$}
\label{sec:sdnunucull}
%

The elements of the $X^{(-)}$ matrix, parametrised by
$c_R^{(-)}, \theta_d^{(-)}, c_I^{(-)}$ enter in the amplitudes for
$s \to d \nu \bar \nu$ and $c \to u \ell^+ \ell^-$ processes.  The
branching ratio for $K \to \pi \nu \bar \nu$ is rather well
determined and probing the SM short-distance contribution.
However, in rare charm meson decays proceeding via
$c \to u \ell^+ \ell^-$ transition, the sensitivity to short distance
SM contributions is reduced due to effective GIM mechanism. In addition, the larger phase space available in $D$ meson decays leads to large long-distance contributions due intermediate kaon and pion rescattering effects. The
ensuing bounds on $z^{(-)}_{\Delta C=1}$ are thus comparably not as constraining as the ones on $z^{(-)}_{\Delta S=1}$ from $s\to
d\nu\bar\nu$. The optimal alignment angle $\theta_d^{(-)*}$ is expected to
be small. In Table~\ref{tab:expInputs} we list the relevant
experimental inputs for $X^{(-)}$.
\begin{table}[!h]
  \centering
  \renewcommand{\arraystretch}{1.1}
  \begin{tabular}{l|c|c}\hline
    Observable & Exp. constraint & Reference\\ \hline \hline
    $\br(K^+ \to \pi^+ \nu \bar \nu)$ & $\left(1.14^{+0.40}_{-0.33}\right)\E{-10}$ & \cite{ParticleDataGroup:2022pth}~(\cite{E949:2008btt,NA62:2021zjw}) \\
    $\br(D^0 \to e^+e^-)$ & $<7.9\E{-8}$ & \cite{Belle:2010ouj}\\
    $\mathcal{B}(D^+ \to \pi^+ e^+ e^-)$ & $< 1.1  \times 10^{-6}$ &  \cite{LHCb:2020car} \\
    $\mathcal{B}(D^0 \to \mu^+ \mu^-) $ &  $ < 3.1  \times 10^{-9}$ & \cite{LHCb:2022uzt}\\              
    $\mathcal{B}(D^+ \to \pi^+ \mu^+ \mu^-)$ & $ < 6.7 \times 10^{-8}$ & \cite{1304.6365} \\
    \hline
    $pp \to e^+ e^-$ & & HighPT~\cite{Allwicher:2022mcg}\\
    $pp \to \mu^+ \mu^-$ & & HighPT~\cite{Allwicher:2022mcg}\\
    $pp \to \tau^+ \tau^-$ & & HighPT~\cite{Allwicher:2022mcg}\\
    \hline
  \end{tabular}
  \caption{Experimental constraints employed as contraints on
    $X^{(-)}$ couplings. Upper bounds are given at {$90\%$ CL}.}
  \label{tab:expInputs}
\end{table}

\subsection{$K^+\to \pi^+ \nu \bar \nu$ and $K_L \to \pi^0 \nu \bar\nu$}

\begin{figure}[!h]
   \begin{center}
  		\includegraphics[scale=1.35]{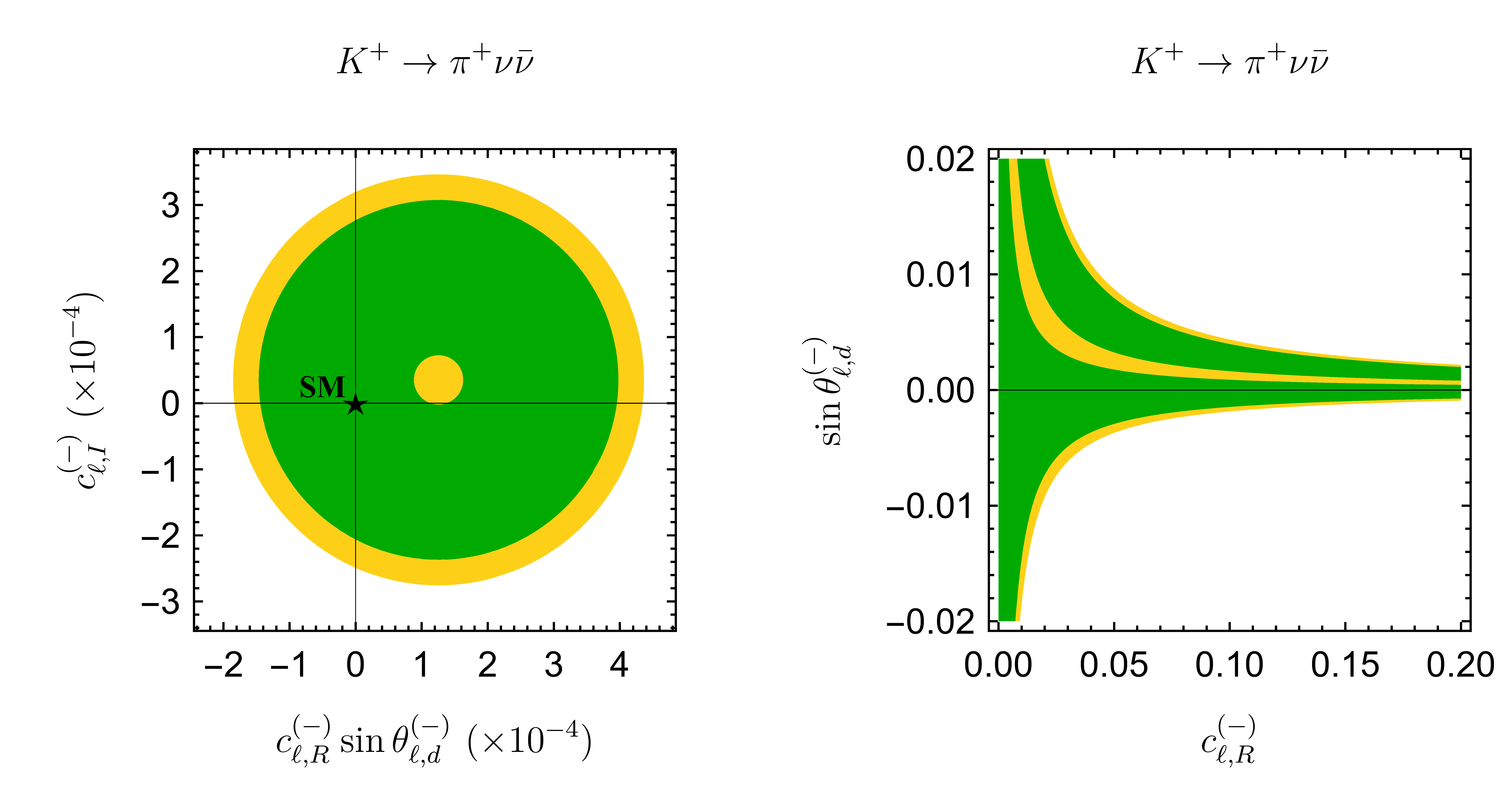}
  \end{center}
  \caption{Left: Constraints in the plane of $c_{\ell,R}^{(-)}
    \sin\theta_{\ell,d}^{(-)}$ and $c_{\ell,I}^{(-)}$ due to the measured
    $\mc{B}(K^+ \to \pi^+ \nu \bar \nu)$. These constraints are valid for
    any lepton flavour $\ell = e,\mu,\tau$. Right: Weakest constraints in the plane of  $c_{\ell,R}^{(-)}$ and $\sin\theta_{\ell,d}^{(-)}$, which occur for $c_{\ell,I}^{(-)} \simeq 0.4 \times 10^{-4}$. }
  \label{fig:Knunu-constraint}
\end{figure}
The differential branching ratio for $ K^\pm \to \pi^\pm  \nu \bar{\nu}$
 can be written as
 \begin{equation}\label{eq:Kpmtopinunudecay}
	\frac{d \mathcal{B}}{d q^2}( K^\pm \to \pi^\pm  \nu \bar{\nu} )=
	(1 + \Delta_{\mathrm{EM}})\frac{G^2_F \alpha_\mrm{em}^2 }{3072 \pi^5 m^3_K \Gamma_{K^\pm}}
	\lambda^{3/2}(m_{\pi}^2, m_{\Delta S=1}^2, q^2)~
	f^2_{+, K\to\pi}(q^2) |C_{L, \nu}^{\Delta S=1}|^2\,,
\end{equation}
where the Wilson coefficient $C_{L,\nu_\ell}$
contains the SM short-distance contribution as well as the contribution of $X^{(-)}$
\begin{align}\label{eq:WETwilsonCoeffCLdown}
  C_{L, \nu}^{\Delta S=1 \rm} = C_{L, \nu}^{\Delta S=1 \rm, SM} + C_{L, \nu_\ell}^{\Delta S=1 \rm, NP}
=\frac{2 (\lambda_c X_c^\ell + \lambda_t X_t)}{\sin^2 \theta_w} +
  \frac{2 \pi v^2}{\alpha_\mrm{em} \Lambda^2} \bigg(c_R^{(-)} \sin
  \theta_d^{(-)} - i c_I^{(-)}\bigg) \,.
\end{align}
The SM contributions have been carefully analysed by several authors~\cite{Brod:2021hsj,Buras:2022wpw,Buras:2022qip,Buras:2021nns,Buras:2015qea}.  Here, $\lambda_q = V_{qs}^* V_{qd}$,  $X_t$ denotes the virtual top-quark
contribution, and $X_c$ the charm-quark contribution. 
The electromagnetic correction $\Delta_{\mathrm{EM}} \simeq -0.3 \%$ was calculated in Ref.~\cite{Mescia:2007kn}. The charm contribution has a mild sensitivity to the lepton flavour, and we take the following values for the Inami-Lim functions
\begin{equation}
X_t = 1.464(17) , \qquad X_c^e \simeq X_c^\mu = 1.04(3) \times 10^{-3}, \qquad X_c^\tau = 0.70(2) \times 10^{-3}\,.
\end{equation}
The uncertainty of $X_t$ has been estimated in Refs.~\cite{Buras:2022wpw,Buras:2005gr, Buras:2006gb}, whereas the uncertainty of ${X_c}$ was recently discussed in Ref.~\cite{Brod:2021hsj}.
For the $K \to \pi$ form factors, we employ lattice
results~\cite{Lubicz:2009ht}. We neglect long-distance
contributions since it has been shown that they are subleading in this decay~\cite{Fajfer:1996tc,Isidori:2005xm}. Current experimental bound on this process is driven by the NA62 measurement~\cite{NA62:2021zjw}, whereas the world average is $\mc{B} (K^+ \to \pi^+ \nu \bar \nu) = (1.14^{+0.40}_{-0.33})\E{-10}$~\cite{ParticleDataGroup:2022pth}.

On the other hand, the amplitude for the $K_L \to \pi^0  \nu \bar{\nu}$ decay is sensitive exclusively to CP-odd effects, leading to the branching fraction
\begin{equation}\label{eq:KLtopinunudecay}
	\frac{d \mathcal{B}}{dq^2} ( K_L \to \pi^0  \nu \bar{\nu}) =
	\frac{G^2_F \alpha_\mrm{em}^2 }{1536 \pi^5 m^3_K \Gamma_{K_L}}
	\lambda^{3/2}(m_{\pi}^2, m_{K}^2, q^2)~
	f^2_{+, K\to\pi}(q^2) \left[\Im(C_{L, \nu}^{\Delta S=1})\right]^2\,,
\end{equation}
where
\begin{align}
\Im({C_{L, \nu}^{\Delta S=1 \rm}})
&=\frac{2\, X_t \im\lambda_t }{\sin^2 \theta_w} - \frac{2 \pi v^2}{\alpha_\mrm{em} \Lambda^2} \,c_{I,\nu}^{(-)}\,.
\end{align}
Using the current 90\% C.L. experimental bound $\mathcal{B}(K_L \to \pi^0 \nu \bar{\nu}) < 3.0 \times 10^{-9}$~\cite{KOTO:2018dsc},  we derive the  constraint $-6.3 \times 10^{-4}  < c^{(-)}_{\ell, I} < 5.6 \times 10^{-4}$, for all $\ell$, which is weaker than the corresponding bound one obtains from $K^+ \to \pi^+ \nu \bar \nu$ (see Fig.~\ref{fig:Knunu-constraint})
\begin{equation}
  -2.8 \times 10^{-4}	<  c^{(-)}_{\ell,I} < 3.5 \times 10^{-4} \,.
  \label{eq:cIminus}
\end{equation}
Note that the bounds on $c^{(-)}_{\tau, I}$ differ
minutely due to $m_\tau$ effects in the loops. 

\subsection{$D^0 \to \ell^+ \ell^-$ and $D^+ \to \pi^+ \ell^+ \ell^-$}

In the SM, the branching ratio for $D^0 \to \mu^+ \mu^-$ is dominated
by long-distance contributions of the $D^0 \to \gamma^* \gamma^*$
intermediate state. However, the relation $\mathcal B (D^0 \to \mu^+
\mu^- ) \approx 2.7\times 10^{-5} \times \mathcal B (D^0 \to \gamma
\gamma)$~\cite{hep-ph/0112235} and the upper bound $\br(D^0 \to \gamma
\gamma) < 8.5\E{-7}$~\cite{Belle:2015pzk} guarantee that the SM
long-distance branching fraction is $\lesssim 10^{-11}$, far below the
current experimental upper bound, $\mathcal{B} ( D^0 \to \mu^+ \mu^{-}
) < 3.1\E{-9}$~\cite{LHCb:2022uzt}. 
Similar conclusion also holds for $D^0 \to e^+e^-$~\cite{hep-ph/0112235,Paul:2010pq,Burdman:2001tf}.
The short-distance contribution to the branching ratio is given by~\cite{Fajfer:2015mia}
\begin{align}
  \mathcal{B}(D^0 \to \ell^+ \ell^-)  &= \frac{G_F^2 \alpha_\mrm{em}^2}{16
                                        \pi^3 \Gamma_{D^0}}  m_D
                                        m_\ell^2 f_D^2 \sqrt{1 - 4
                                        m_\ell^2/m_D^2} \,  |C^{\Delta C=1 \rm}_{10}|^2 \,,
\end{align}
where we can safely neglect SM contributions to $C^{\Delta C=1 \rm}_{10}$.
On the other hand, the differential branching ratio of $D \to \pi \ell^+ \ell^-$ is sensitive to both vector $C_9^{\Delta C=1}$ and axial $C_{10}^{\Delta C=1}$ Wilson coefficients
	\begin{equation}\label{eq:Dtopiellell}
		\frac{d\br}{dq^2} ( D^+ \to \pi^+  \ell^+ \ell^- )  =
		\frac{G^2_F \alpha_\mrm{em}^2}{1536 \pi^5 m^3_D \Gamma_{D^0}}
		\lambda^{1/2}(m_{\pi}^2, m_D^2, q^2) \sqrt{1 - 4 m_\ell^2/q^2} ~
		\Big(\mathcal{I}^D_V(q^2) + \mathcal{I}^D_A(q^2)\Big),
	\end{equation}
where axial and vector lepton current contributions are
\begin{align}
\mathcal{I}^D_A(q^2) &=  |C^{\Delta C=1 \rm}_{10}|^2\Bigg( f^2_+(q^2) \lambda(q^2, m_D^2, m_\pi^2) \Big(1 - \frac{4 m_\ell^2}{q^2} \Big) +  6\frac{m_\ell^2}{q^2}(m_D^2 - m_\pi^2)^2 f_0^2(q^2) \Bigg), \nonumber\\
\mathcal{I}^D_V(q^2) &= |C^{\Delta C=1 \rm}_{9}|^2 f^2_+(q^2) \lambda(q^2, m_D^2, m_\pi^2) \Big(1 + \frac{2 m_\ell^2}{q^2} \Big)  + \ldots \,,
\label{eq:ID}
\end{align}
and $\lambda(x,y,z) = (x+y+z)^2-4(xy+yz+zx)$.
Our results coincide with the ones given in Ref.~\cite{Mescia:2007kn}.
The expression \eqref{eq:Dtopiellell} can be employed also for $D^0 \to \pi^0 \ell^+ \ell^-$ albeit with an additional factor $1/2$.
The limit extraction on short-distance NP for $D^+ \to \pi^+ \ell^+ \ell^-$ is more complicated due to diverse resonant long-distance contributions (indicated by ellipsis in Eq.~\eqref{eq:ID}). One solution is to integrate over the high-$q^2$ phase-space portion which is free of those contributions (if appropriate bounds exist, like e.g. for $D^+ \to \pi^+ \mu^+ \mu^-$ \cite{1304.6365}), or integrating over the whole kinematic region together with parametrizing the dominant long-distance contributions in terms of Breit-Wigner resonances~\cite{1909.11108}. A careful analysis taking the latter approach was done in Ref.~\cite{2003.12421}, from where we take the resulting limits on $|C_{9,10}^{\Delta C=1}|$.\\
We have checked that the limit from $D^0 \to \mu^+ \mu^-$
branching fraction is stronger than the one coming from $D^+ \to \pi^+
\mu^+ \mu^-$, whereas the opposite is true for modes with electrons. 
 Explicitly,  in the case of muons, they read
\begin{equation}\label{eq:LimforDPiMuMu}
	\Big|
	c^{(-)}_{\mu,R} \sin (\theta_{\mu,d}^{(-)} - 2 \theta_c) - i c_{\mu,I}^{(-)} 
	\Big| 
	< 2.4 \times 10^{-2}\,.
\end{equation}
Due to the limits coming from the branching ratio for $K^+ \to \pi^+ \nu \bar{\nu}$ that are $\sim 10^{-4}$ for both $c_I^{(-)}$ and $c_R^{(-)} \sin \theta_d^{(-)}$, see Fig.~\ref{fig:Knunu-constraint}, 
we can simplify Eq.~(\ref{eq:LimforDPiMuMu}) as 
\begin{equation}
\Big|\,c_{\mu,R}^{(-)} \cos \theta_{\mu,d}^{(-)}\,\Big| < \frac{2.4 \times 10^{-2}}{\sin 2 \theta_c}  = 0.055\,.
\end{equation}
Similarly, for electrons we find
\begin{equation}
	\Big|\,c_{e,R}^{(-)} \cos \theta_{e,d}^{(-)}\,\Big| <  0.43\,.
\end{equation}

%
\section{$s \to d \ell^+ \ell^-$ and $c \to u \nu \bar \nu$}
\label{sec:sdllcununu}
%

The elements of the $X^{(+)}$ matrix enter in the amplitudes for
$s \to d  \ell^+ \ell^-$ and $c \to u \nu \bar \nu$ processes. In the charm sector, there exists a single bound on the branching
ratio  $\mathcal B(D \to \pi \nu \bar \nu)$ obtained recently by BESS III, which is still well above the GIM suppressed SM prediction. In
contrast there already exist numerous experimental probes of the strange
sector -- the branching ratios  $\mathcal B (K_{L/S} \to \ell^+ \ell^-)$, as well as
semileptonic decays i.e. $K_L \to \pi^0 \ell^+ \ell^-$. The kaon observables
with charged leptons however receive sizable long-distance non-perturbative SM
contributions and thus suffer from larger theoretical uncertainties. This renders the
HighPT constraints on the $(+)$ sector to be even more important than
in the case of $(-)$ couplings.

\begin{table}[!h]
	\centering
	\renewcommand{\arraystretch}{1.1}
	\begin{tabular}{l|c|c}\hline 
		Observable & Exp. constraint & Reference\\\hline  \hline 
		$\br(K_L \to e^+ e^-)$ & $\left(9^{+6}_{-4}\right)\E{-12}$ & \cite{ParticleDataGroup:2022pth}~(\cite{BNLE871:1998bii}) \\
		$\br(K_L \to \mu^+\mu^-)$ & $(6.84 \pm 0.11)\E{-9}$ & \cite{ParticleDataGroup:2022pth}~(\cite{E871:2000wvm})\\
		$\br(K_L \to \pi^0 e^+ e^-)$ & $ < 2.8 \E{-10}$ & \cite{ParticleDataGroup:2022pth}\\
		$\br(K_L \to \pi^0 \mu^+\mu^-)$ & $ < 3.8 \E{-10}$ & \cite{ParticleDataGroup:2022pth}~(\cite{KTEV:2000ngj})\\
		$\br(D^0 \to \pi^0 \nu \bar{\nu})$ & $<2.1\E{-4}$ & \cite{BESIII:2021slf}\\
		\hline
		$pp \to e^+ e^-$ & & HighPT~\cite{Allwicher:2022mcg}\\
		$pp \to \mu^+ \mu^-$ & & HighPT~\cite{Allwicher:2022mcg}\\
		$pp \to \tau^+ \tau^-$ & & HighPT~\cite{Allwicher:2022mcg}\\
		\hline
	\end{tabular}
	\caption{Experimental constraints employed as contraints on
		$X^{(+)}$ couplings. Upper bounds are given at $90 \%$ CL.}
	\label{tab:expInputs2}
\end{table}

\subsection{$K_L \to \pi^0 \ell^+ \ell^-$ and $K_L \to \ell^+ \ell^-$}
The rare semileptonic decay $K_L \to \pi^0 \ell^+ \ell^-$ is
  sensitive to CP-odd short distance effects, parameterized by
  $c_I^{(+)}$. However, the SM amplitude is
dominated by long-distance dynamics. One has contributions from
indirect CPV ($K_L \to K_S$ transition followed by
$K_S \to \pi^0 \ell^+ \ell^-$), as well as CP-conserving long-distance
$K_L \to \pi^0 + 2 \gamma^* \to \pi^0 \ell^+
\ell^-$~\cite{DAmbrosio:1998gur,Buchalla:2003sj,Isidori:2004rb}.
Within our framework, the short-distance contribution of NP to vector and axial-vector coefficients is of the form~\cite{Mescia:2006jd}:
\begin{align}
\mathcal{B}(K_L \to \pi^0 \mu^+ \mu^-) &= \big(1.09 \, (w_{7V}^2 + 2.32 w_{7A}^2) \pm 2.63 \, w_{7V} |a_S| + 3.36 \, |a_S|^2 + 5.2 \big) \times 10^{-12}\,, \label{KtoPimumudecay}\\
\mathcal{B}(K_L \to \pi^0 e^+ e^-) &= \big(4.62 \, (w_{7V}^2 + w_{7A}^2) \pm 11.3 \, w_{7V} |a_S| + 14.5 \, |a_S|^2 \big) \times 10^{-12}\,, \label{KtoPieedecay}
\end{align}
with:
\begin{align}
w_{7V} &= y_{7V} - \frac{v^2}{2 \, \alpha_\mrm{em}  \Lambda^2\,\im \lambda_t }c^{(+)}_I  \,,\\
w_{7A} &= y_{7A}  + \frac{v^2}{2 \, \alpha_\mrm{em}  \Lambda^2 \im \lambda_t}c^{(+)}_I  \,.
\end{align}
Here $y_{7V} = 0.735$, $y_{7A} = -0.700$~\cite{hep-ph/9512380}, $|a_S| = 1.20\pm 0.20$~\cite{NA481:2003cfm}.
Notice the ambiguity due to the unknown sign of the interference term between $w_{7V}$ and $|a_S|$.
Experimental bounds for both lepton flavours 
\begin{equation}
	\mathcal{B}(K_L \to \pi^0 \mu^+ \mu^-)  < 3.8 \times 10^{-10}\,\text{\cite{KTEV:2000ngj}},\qquad
	\mathcal{B}(K_L \to \pi^0 e^+ e^-) < 2.8 \times 10^{-10}\,\text{\cite{ParticleDataGroup:2022pth}},
      \end{equation}
are given at 90\% CL and are an order of magnitude above the SM
prediction, as can be seen when we comparing them to the respective SM predictions:
\begin{align}
  \mathcal{B}(K_L \to \pi^0 \mu^+ \mu^-)_{\rm SM} = \left\{\begin{array}{lcl} (1.41^{+0.28}_{-0.26})  \times 10^{-11}& ; & +\textrm{ sign}\\
      (0.95^{+0.22}_{-0.21}) \times 10^{-11} & ; & -\textrm{ sign}\end{array}\right. \,, \\
    \mathcal{B}(K_L \to \pi^0 e^+ e^-)_{\rm SM} = \left\{\begin{array}{lcl} (3.54^{+0.98}_{-0.85})  \times 10^{-11}& ; & +\textrm{ sign}\\
  	(1.56^{+0.62}_{-0.49}) \times 10^{-11} & ; & -\textrm{ sign}\end{array}\right. \,.
\end{align}
The $+$ and $-$ signs correspond to the sign chosen in the
interference term in Eqs.~\eqref{KtoPimumudecay}
and~\eqref{KtoPieedecay}. Since the SM prediction is an order of
magnitude below the current experimental limits, we approximate the
likelihood by neglecting the SM contributions in the fit. This enables
us to derive  directly the constraints\footnote{In addition, we have checked explicitly using results of Ref.~\cite{Mescia:2006jd}, that the limits from $K_S \to \ell^+ \ell^-$ are weaker compared to $K_L \to \pi \ell\ell$.}
\begin{equation}
  \label{eq:CIplus}
  \big|\im[c_{e,I}^{(+)}]\big| < 2 \times 10^{-4}, \qquad \big|\im[c_{\mu,I}^{(+)}]\big| < 4 \times 10^{-4} .
\end{equation}

\subsection{$K_L \to \ell^+ \ell^-$}

This decay mode is also dominated by long-distance SM contributions.
We explore the information coming from the $K_L \to \mu^+ \mu^-$ and $K_L \to e^+ e^-$ decays as already considered in Refs.~\cite{Mescia:2006jd, Geng:2021fog} (see also \cite{Dery:2021mct} for explicit analytic expressions). 
In our analysis we employ the lattice QCD result for the decay constant $\langle 0| \bar s \gamma^\mu \gamma^5 d | K^0(p) \rangle = i f_K p^\mu$  with $f_K = 0.1557 \,\mathrm{GeV}$~\cite{Aoki:2021kgd}.
The branching ratio is then given by
\begin{equation}\
\mathcal{B}(K_L \to \ell^+ \ell^-) = \frac{G_F^2 \alpha_\mrm{em}^2}{8\pi^3} f_{K}^2 m_K m_\ell^2 \sqrt{1 - 4 m_\ell^2 / m_K^2} \left|C_{10}^{\Delta S=1 \rm}\right|^2 \,,
\end{equation}
where
\begin{equation}
  C_{10}^{\Delta S=1 \rm} = C_{10}^{\Delta S=1 \rm, SM} + C_{10}^{\Delta S=1 \rm, NP} = -2\pi\left(\re(\lambda_t y_{7A}) + \re(\lambda_c y_c) - \frac{v^2}{ 2\alpha_\mrm{em} \Lambda^2} c_R^{(+)} \sin \theta_d^{(+)}\right) + \frac{A^\ell_{L \gamma \gamma}}{\sin^2 \theta_W} \,.
\end{equation}
Notice that this process is sensitive only to CP conserving parameters
$c_R^{(+)}$ and $\theta_d^{(+)}$. The factor of $2\pi$ comes from the
different conventions for the effective Hamiltonian relative
to~\cite{Mescia:2006jd}. Here, the long-distance two-photon
intermediate state contribution has a relative sign ambiguity and is currently estimated as~\cite{hep-ph/0311084,hep-ph/0508189}:
\begin{align}
|A^\mu_{L \gamma \gamma}| &=  1.98 \times 10^{-4} \big(0.71 \pm 0.15 \pm 1.0 - 5.21\,i  \big)\,,\\
|A^e_{L \gamma \gamma}| &=   1.98 \times 10^{-4} \big(31.91 \pm 0.22
                        \pm 1.0 - 21.61\, i  \big)\,.
\end{align}
The resulting SM predictions for the branching fractions are
 $\mc{B}(K_L \to \mu^+ \mu^-)=(7.64 \pm 0.73) \times 10^{-9}$~\cite{Geng:2021fog}, and $\mc{B}(K_L \to e^+ e^-)=(9.0 \pm 0.5) \times 10^{-12}$~\cite{hep-ph/9711377}.
Both theoretical estimates are comparable to the experimentally measured values~\cite{ParticleDataGroup:2022pth}:
\begin{align}
	\mathcal{B}(K_L \to \mu^+ \mu^-)^{\rm exp} &= (6.84 \pm 0.11) \times 10^{-9} \,,\\
	\mathcal{B}(K_L \to e^+ e^-)^{\rm exp} &= 9^{+6}_{-4}  \times 10^{-12}\,.
\end{align}

\subsection{$D^0 \to \pi^0 \nu \bar{\nu}$}
\label{sec:dpinunu}
The differential branching fraction of this decay mode depends only on the $C_{L, \nu}^{\Delta C=1}$ coefficient:
\begin{equation}
  \label{dtopidecay}
	\frac{d \mathcal{B}}{dq^2} ( D^0 \to \pi^0  \nu \bar{\nu})  =
	\frac{G^2_F \alpha_\mrm{em}^2 }{3072 \pi^5 m^3_D \Gamma_{D^0}}
	\lambda^{3/2}(m_{\pi}^2, m_D^2, q^2) ~
	f^2_{+, D \to \pi}(q^2)\left|C_{L, \nu}^{\Delta C=1} \right|^{2}\,.
\end{equation}  
The above expression corresponds to a final state with a specific
neutrino flavour in the final state. Measurable branching fraction is
obtained by summing over all three neutrino
flavours~\cite{Bause:2020xzj}. In our setup exactly one Wilson
coefficient $C_{L,\nu}^\mrm{up}$ contains a BSM contribution, while
remaining flavours are purely SM. The form factor $f_+(q^2)$ in
Eq.~\eqref{dtopidecay} is obtained from the form factor of charged
transition $D^+ \to \pi^+$, scaling it by $1 / \sqrt{2}$ due to the
isospin wavefunction of the $\pi^0$ state. We employ the form factor
obtained using lattice QCD~\cite{1706.03017}. On the
experimental side, there is a single upper limit result due to BESSIII,
$\mc{B}(D^0 \to \pi^0 \nu \bar \nu) < 2.1\E{-4}$ at 90\%
CL~\cite{BESIII:2021slf}. This bound currently results in a relatively weak
constraint that cuts away large values of $|c_{\ell,R}^{(+)}
\sin(\theta_{\ell,d}^{(+)} -2\theta_c)|\lesssim 3$.
It is however not competitive with the charged current constraints from LHC high-$p_T$ tails and
measurements of CKM elements, shown in Fig.~\ref{fig:cPlusLimits2}.

%
\section{High-$p_T$ limits}
\label{sec:HighPTlimits}
%

An important set of limits  arise when one confronts the measured cross sections of $p p\to \ell^+ \ell^-$ at
the LHC against the theoretical predictions in the SM complemented by
neutral current effective interactions in Eq.~\eqref{eq:FCNCs}. Extracting
the bounds from high-energy $pp \to \ell^+ \ell^-$ processes is more
involved due to contributions both from the up- and
down-quarks. Summing over quark flavors found in the proton, we have
the following set of interactions contributing incoherently
to the neutral-current cross section:
\begin{align}\label{eq:highPToperators}
	\sigma_{\mathrm{high}-p_{T}} 
	&\supset 2 \int_{\tau_{\rm min}}^{\tau_{\rm max}} d \tau \frac{\tau \, S_{\rm had}}{144 \pi} \times \nonumber \\
	\Big[&\mathcal{L}_{u \bar{u}}  \left(\lambda^{(-)} - c_R^{(-)} \cos ( 2 \theta_c - \theta_d^{(-)}) + F_{\mathrm{SM}, u\bar{u}}\right)^2 + \mathcal{L}_{c \bar{c}}  \left(\lambda^{(-)} + c_R^{(-)} \cos (2 \theta_c - \theta_d^{(-)}) + F_{\mathrm{SM},u\bar{u}}\right)^2 \nonumber \\ 
	+ &\mathcal{L}_{d \bar{d}} \left(\lambda^{(+)} - c_R^{(+)} \cos \theta_d^{(+)} + F_{\mathrm{SM},d\bar{d}}\right)^2 + \mathcal{L}_{s \bar{s}} \left(\lambda^{(+)} + c_R^{(+)} \cos \theta_d^{(+)} + F_{\mathrm{SM},d\bar{d}}\right)^2 \\
	+ &(\mathcal{L}_{u \bar{c}}  +  \mathcal{L}_{c \bar{u}}) \left( \big(c_R^{(-)} \sin (2 \theta_c - \theta_d^{(-)})\big)^2 + (c_I^{(-)})^2 \right) + (\mathcal{L}_{d \bar{s}}  +  \mathcal{L}_{s \bar{d}}) \left(\big(c_R^{(+)} \sin \theta_d^{(+)}\big)^2 + (c_I^{(+)})^2 \right)\Big], \nonumber
\end{align}

\noindent together with other interactions which are unaffected by the
SMEFT operators. Here, $F_{\mathrm{SM},q\bar{q}}$ denote the
corresponding SM contributions, which are due to the $\gamma$ and $Z$
$s$-channel processes
\begin{align}
 F_{\mathrm{SM}, u\bar{u}}\, (p^2) = \frac{Q_u e^2}{p^2} + \frac{g_{Z, u\bar{u}}\, g_{Z, \ell \bar{\ell}}}{p^2 - m_Z^2 + i m_Z \Gamma_Z}\,, \nonumber \\
 F_{\mathrm{SM}, d\bar{d}}\, (p^2) = \frac{Q_d e^2}{p^2} + \frac{g_{Z, d\bar{d}}\, g_{Z, \ell \bar{\ell}}}{p^2 - m_Z^2 + i m_Z \Gamma_Z}\,,
\end{align} 
where $g_{Z,f\bar f} = \frac{2 m_Z}{v} (T_f^3 - Q_f \sin^2 \theta_W)$ are the $Z$ couplings to the fermions~\cite{Greljo:2017vvb}
and $p^2 = \hat{s} =  \tau\, s_{\rm had}$, $\sqrt{s_{\rm had}} = 13 \, \mathrm{TeV}$. Finally, $\mathcal{L}_{q\bar{q}}$ denotes the luminosity function
\begin{equation}
\mathcal{L}_{q\bar{q}} \equiv \mathcal{L}_{q \bar{q}} (\tau, \mu_F) = \int_{\tau}^{1} \frac{dx}{x} f_q (x, \mu_F) f_{\bar{q}} (\tau/x, \mu_F) \,.
\end{equation} Note that we ignore SM contributions to the FCNCs,
since they are suppressed by a loop factor and GIM.
The later is very effective at high energies resulting in
negligible SM effects on the cross section. We employ the HighPT
package~\cite{2207.10756, 2207.10714} based on ATLAS~\cite{ATLAS:2020zms} and
CMS~\cite{CMS:2021ctt} measurements of $pp\to \ell\ell$, in order to find bounds from high-$p_T$ LHC data on our SMEFT parameter space.

It is evident from Eq.~\eqref{eq:highPToperators} that the weakest bounds
on $c_R^{(-)},\theta_d^{(-)}$   will occur if we set $c_R^{(+)} =
c_I^{(+)} = 0$. Furthermore, our bounds are derived by marginalizing
over the trace parameters $\lambda^{(\pm)}$. The results of the
marginalization procedure  can be understood in advance; $\lambda^{(+)}$ will pick up a value such that $\int d\tau \,\tau \Big( \mathcal{L}_{d \bar{d}} \big(\lambda^{(+)} + F_{\mathrm{SM}, d\bar{d}}\big)^2 + \mathcal{L}_{s \bar{s}}\big(\lambda^{(+)} + F_{\mathrm{SM}, d\bar{d}}\big)^2 \Big) = 0$. 
On the other hand, $\lambda^{(-)}$ needs to be $\lambda^{(-)} \sim
c_R^{(-)} \cos 2\theta_c$ to reduce the dominant $u\bar{u}$
contribution, however, the limit will eventually be saturated through
the $c \bar{c}$, $u \bar{c}$ and $\bar{u} c$ initial states. The same
arguments apply if we are interested in the weakest bounds on
$c_R^{(+)}, \theta_d^{(+)}$.

The bounds in both sectors are correlated, so setting a
non-zero value to $c_R^{(-)}, \theta_d^{(-)}$ will shrink the allowed
space in the plus region, and vice versa.  Indeed the bounds will be applied separately
for the $(-)$ and the $(+)$ sector assuming that couplings $(\pm)$ are
zero when deriving bounds on $(\mp)$.

%
\section{Charged currents at low energies and at high-$p_T$}
\label{sec:CC}
%
In the previous sections we have studied  $c^{(+)}$ and
$c^{(-)}$ separately and independently of each other. Such approach is strictly valid
only if $c^{(+)} = c^{(-)}$ and there is no further effect in charged
currents, see Eq.~\eqref{eq:CCs}.  The Lagrangian that governs the
semileptonic charged-current contributions at low energies can be written as:
\begin{equation}\label{eq:chargedcurrentL}
 \mathcal{L}_\mrm{cc} =  \frac{G_F}{\sqrt{2}} V^{\ell}_{ij} (\bar{u}_i \gamma_{\mu}(1 - \gamma_5) d_j)(\bar{\ell} \gamma^{\mu}(1 - \gamma_5) \nu_\ell) + \mrm{h.c.},
\end{equation}
where $V^{\ell}_{ij} = V_{ij} + \Delta V_{ij}^{(\ell)}$ represents
the NP modified effective CKM coupling. Individual lepton-specific CKM modifications
depend on $c_{R,I}^{(\pm)}$, $\theta_d^{(\pm)}$, $\theta_c$ as well as
on the trace parameters $\lambda$ as can be seen in Eq.~\eqref{eq:CKMmods}.
These modifications are subject to strong constraints from tree-level probes of CKM elements, such as superallowed $\beta$ decays, and semileptonic $K$, $D$ and $\tau$ decays.
Note that we can completely remove the dependence on the trace
parameter $\lambda^{(3,\ell)}$ by considering the following
combinations of effective CKM elements:
\begin{subequations} 
  \label{eq:CKMclean}
  \begin{align}
  V^{\ell}_{us} + V^{\ell}_{cd} &= \frac{v^2}{\Lambda^2}  \left[ \cx{+}_R \sin \left(\theta_c + \theta_d^{(+)}\right) - \cx{-}_R \sin\left(\theta_c + \theta_d^{(-)}\right)\right],\\
    V^{\ell}_{ud} - V^{\ell}_{cs} &= \frac{v^2}{\Lambda^2}  \left[ -\cx{+}_R \cos \left(\theta_c + \theta_d^{(+)}\right) + \cx{-}_R \cos\left(\theta_c + \theta_d^{(-)}\right) + i \left(\cx{+}_I - \cx{-}_I\right) \sin\theta_c \right].
\end{align}
\end{subequations}
In the following we omit the $c_I$ terms since they do not interfere
with the SM and their size is severely constrained by neutral current
processes. By squaring and summing  Eqs.~\eqref{eq:CKMclean} we
can even eliminate the dependence on the Cabibbo angle:
\begin{equation}
  \label{eq:CKMCabibboFree}
  \left(V^{\ell}_{us} + V^{\ell}_{cd}\right)^2 +   \left(V^{\ell}_{cd} - V^{\ell}_{cs}\right)^2 = \left(\frac{v}{\Lambda}\right)^4 \left[\left(\cx{+}_R\right)^2 + \left(\cx{-}_R\right)^2 -2 \cx{+}_R \cx{-}_R \cos\left(\theta_d^{(+)} - \theta_d^{(-)} \right) \right]\,.
\end{equation}
For completeness we also state the remaining two combinations
\begin{equation} 
  \label{eq:CKMtrace}
  \begin{pmatrix}
    V^{\ell}_{us} - V^{\ell}_{cd} \\
    V^{\ell}_{ud} + V^{\ell}_{cs} 
  \end{pmatrix}
  =
  2 \left[ 1+\frac{v^2}{\Lambda^2} \lambda^{(\ell)}\right]
  \begin{pmatrix}
    \sin \theta_c\\ \cos\theta_c
  \end{pmatrix}\,.
\end{equation}
These relations are free of neutral current parameters ($c_R$, $c_I$,
$\theta_d$) and can be inverted to determine the Cabibbo angle and the
trace parameter:
\begin{equation}
  \tan \theta_c = \frac{V^{\ell}_{us} - V^{\ell}_{cd}}{V^{\ell}_{ud} + V^{\ell}_{cs}}\,,\qquad \frac{v^2}{\Lambda^2} \lambda^{(3,\ell)} \approx \frac{(V^{\ell}_{us} - V^{\ell}_{cd})^2 + (V^{\ell}_{ud} + V^{\ell}_{cs})^2-4}{8}\,.
\end{equation}
Experimental information on $V_{ij}^\ell$ has to be extracted from
lepton specific processes. We will impose as experimental constraints
super-allowed $\beta$ decay, charged pion, kaon, $\tau$ and charm
decays. We detail the experimental inputs of
charged-current processes and the extraction procedure in
Appendix~\ref{app:CKMextraction}.

\begin{figure}[!h]
	\centering
	\includegraphics[scale=1.]{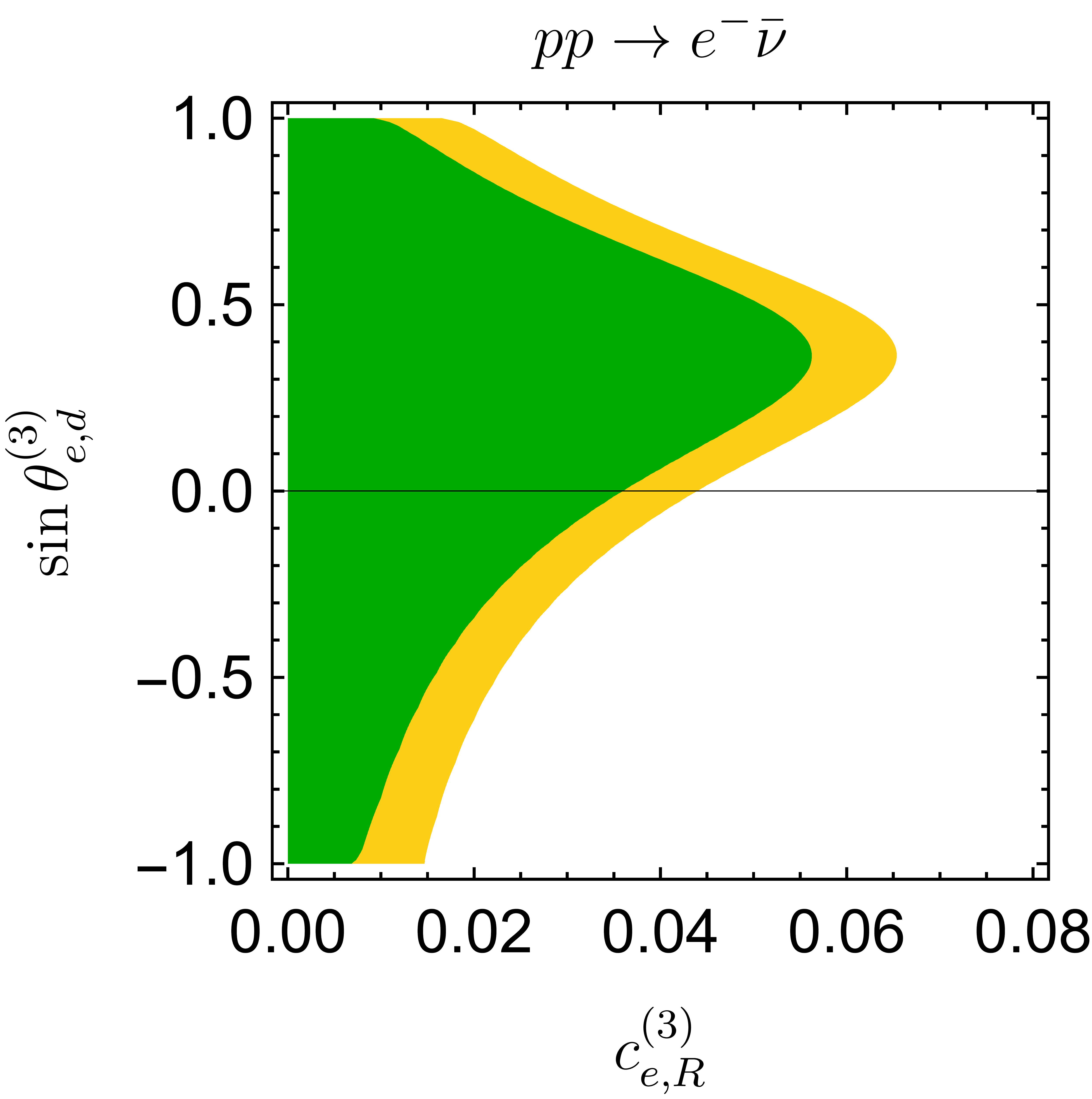}\\
	\includegraphics[scale=1.]{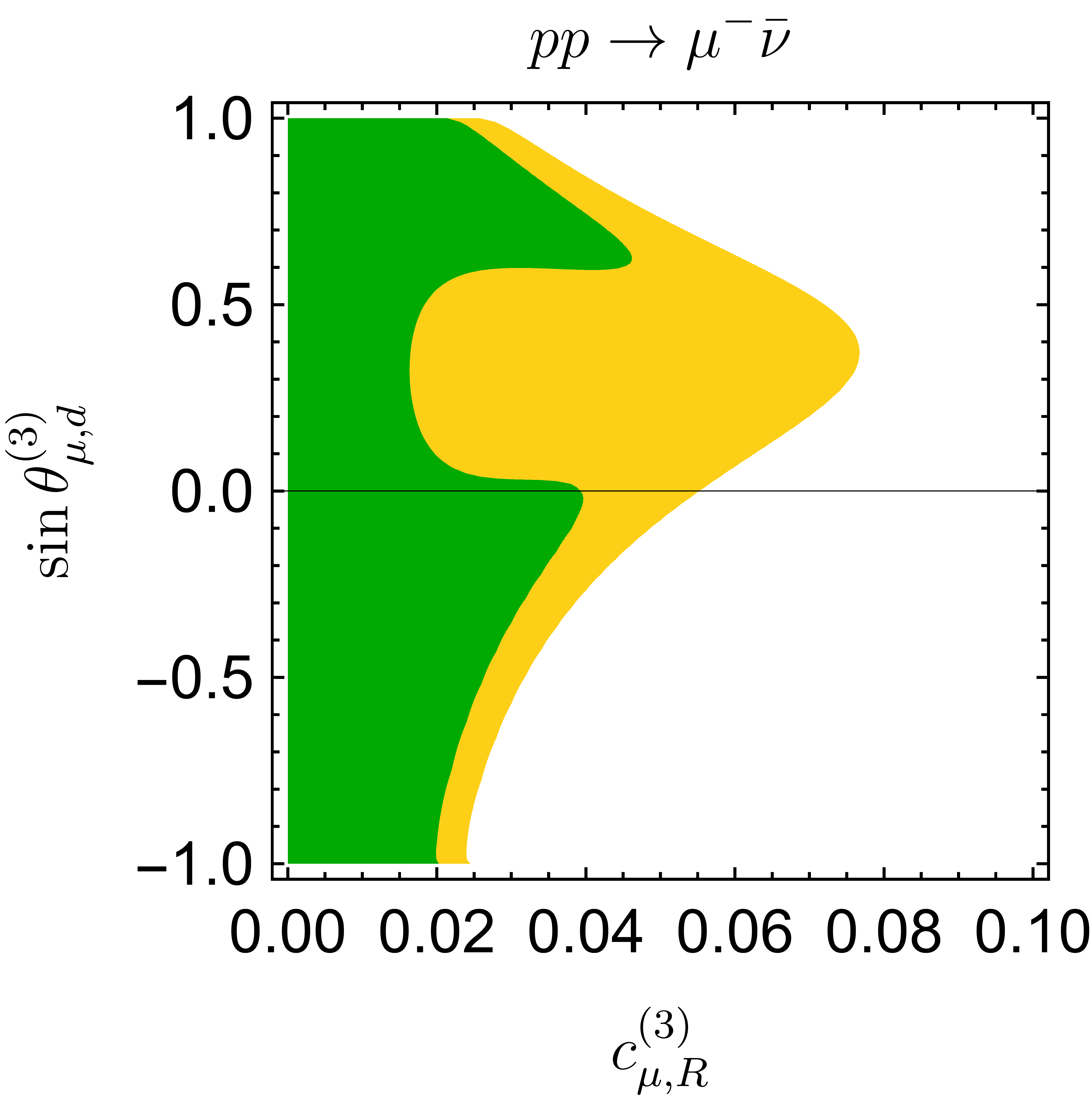}\\
	\includegraphics[scale=1.]{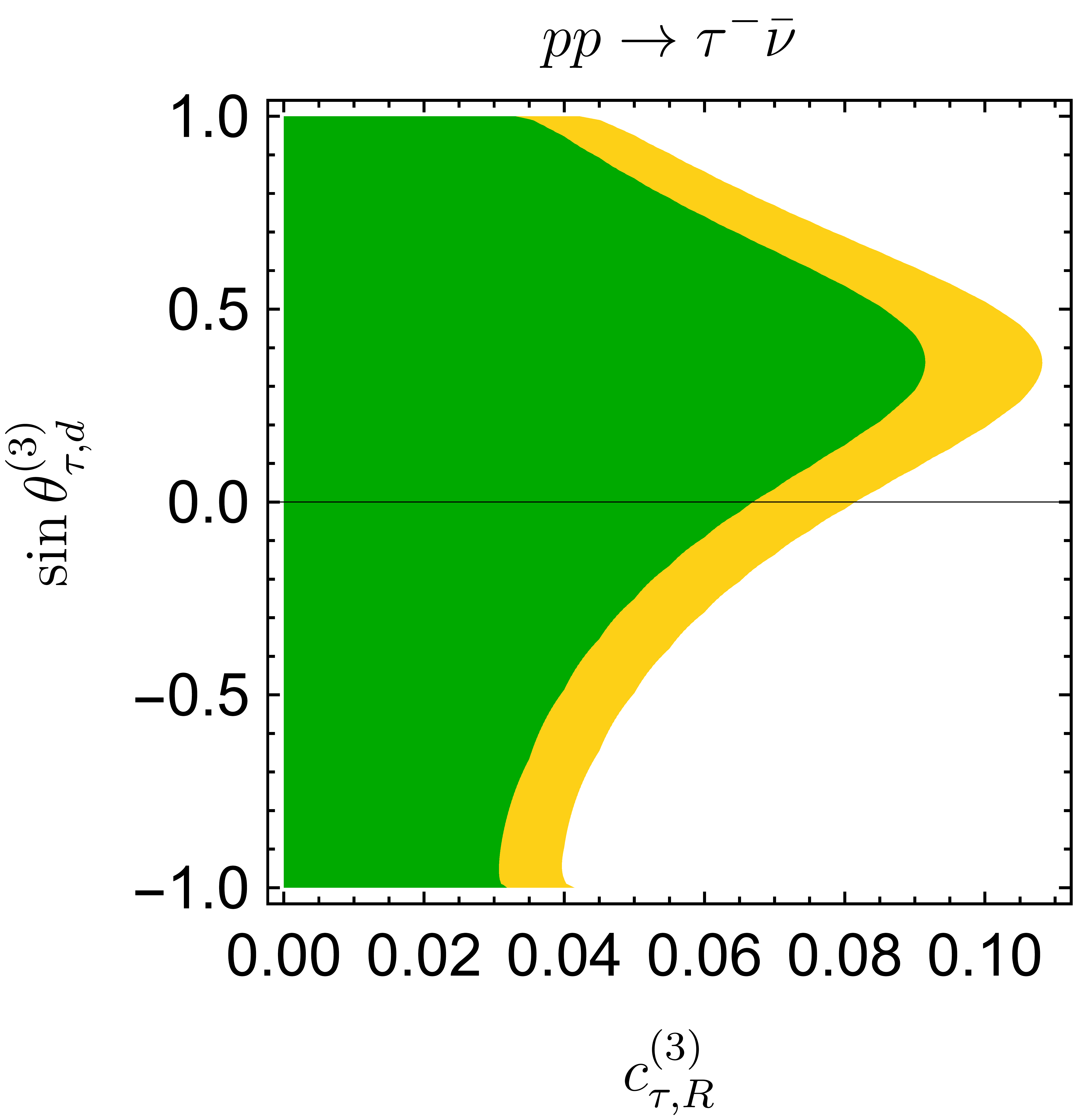}
	\caption{Bounds on the triplet operators from high-$p_T$ charged current processes~\cite{Allwicher:2022mcg}. See text for details.}
	\label{fig:chargedHighPTbounds}
\end{figure}

As for the high-$p_T$ constraints, analogous expressions to
Eq.~\eqref{eq:highPToperators} hold for charged currents processes
($pp \to \ell \nu$) which bound the parameter space only in the
$c^{(3)} = (c^{(-)} - c^{(+)})/2$ direction (see Fig.~\ref{fig:chargedHighPTbounds}). Since the neutral current
constraints allow for larger effects in $c^{(+)}$ than in
$c^{(-)}$ the charged current constraints are relevant only for
$c^{(+)}$.

%
\section{Results}
\label{sec:results}
%

\begin{figure}[!h]
  \centering\begin{tabular}{l}
    \includegraphics[scale=0.94]{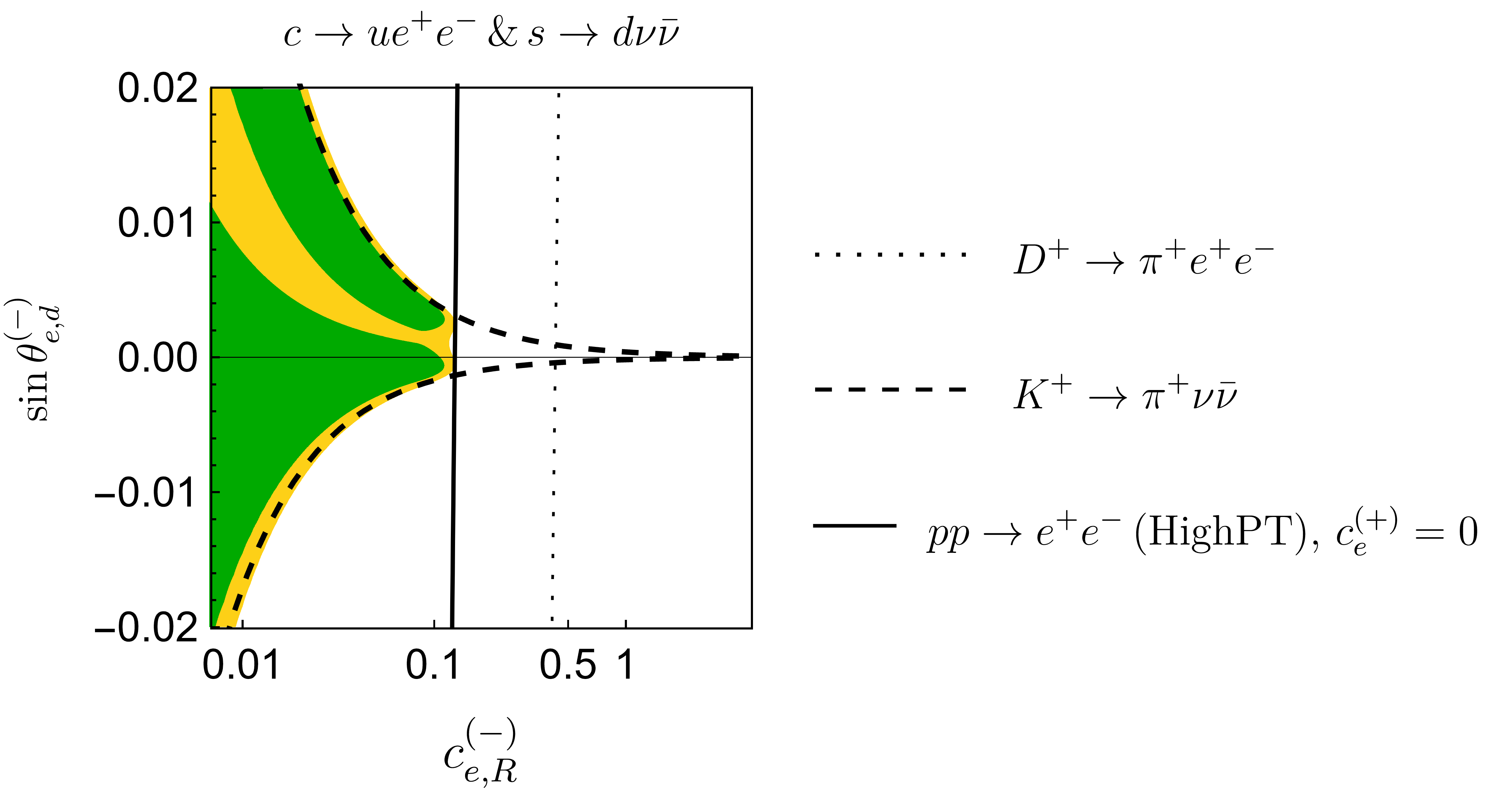}\\
    \includegraphics[scale=0.97]{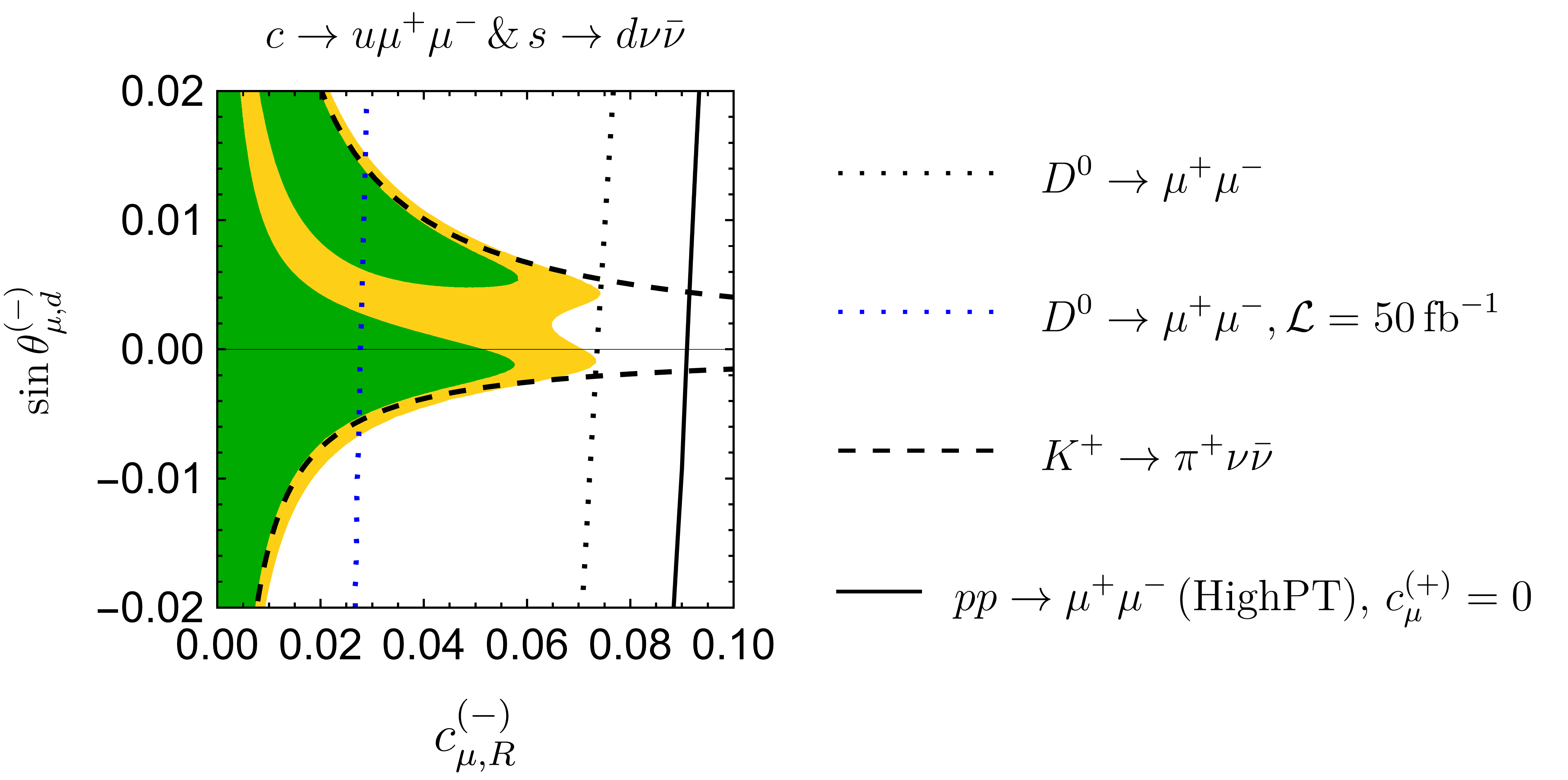}\\
    \includegraphics[scale=0.94]{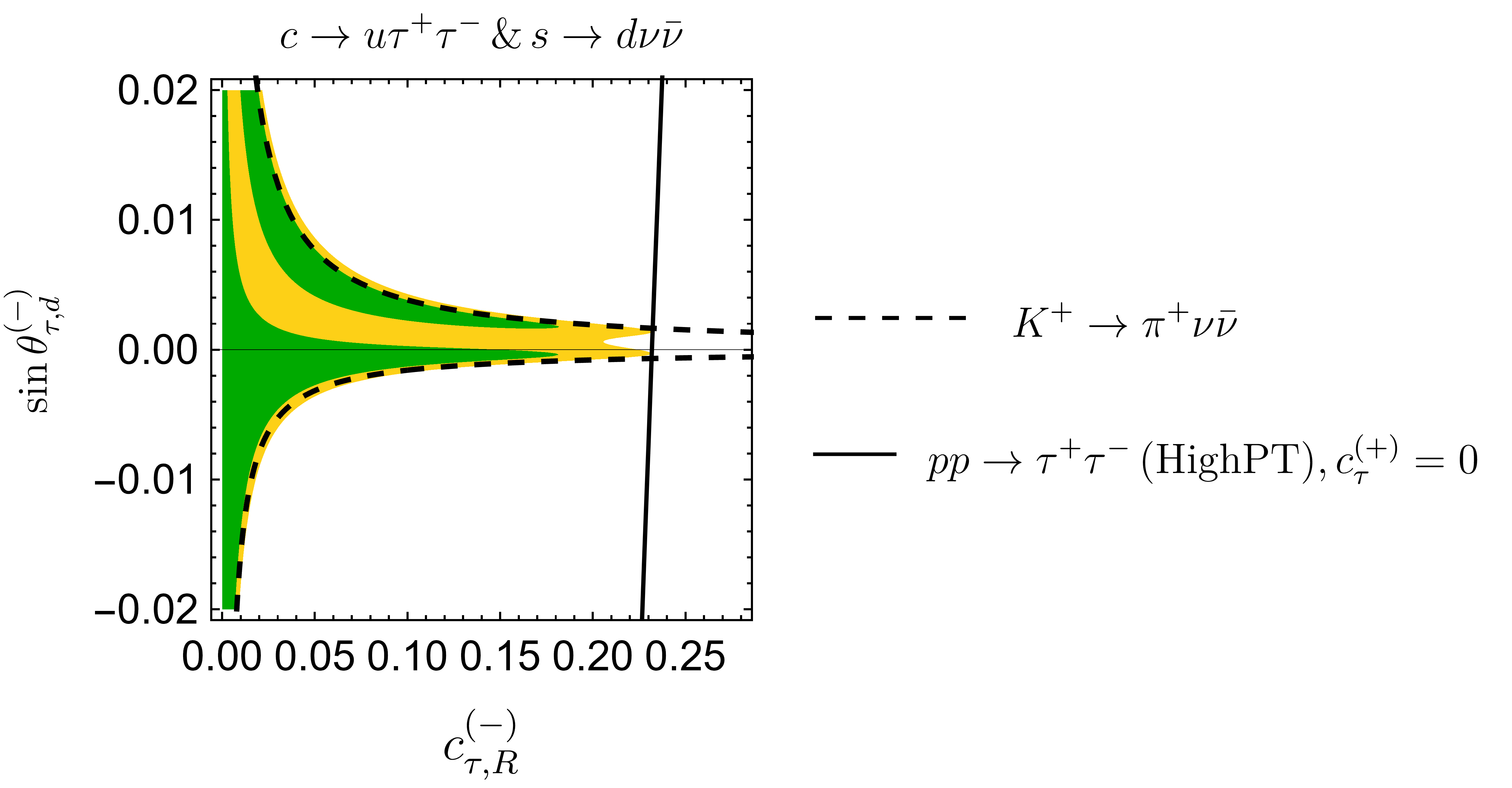}\\
   \end{tabular}
  \caption{Bounds on the $(-)$ operators. The imaginary part (for all
    three lepton generations) was taken to be $c_{\ell, I}^{(-)} \simeq 0.4 \times 10^{-4}$, for which the bounds on
    $c^{(-)}_{\ell,R} \sin \theta^{(-)}_{\ell,d}$ from $K^+ \to \pi^+ \nu
    \bar \nu$ are the weakest. The green and yellow shaded regions correspond to $68\%$ CL and $95\%$ CL allowed regions of the global fit. The HighPT bounds were derived under the assumption that $c^{(+)}_{\ell, R} = c^{(+)}_{\ell, I} = 0$.}
  \label{fig:cRtanThetaMinus}
\end{figure}

\begin{figure}[!h]
  \centering
  \begin{tabular}{l}
    \includegraphics[scale=1.29]{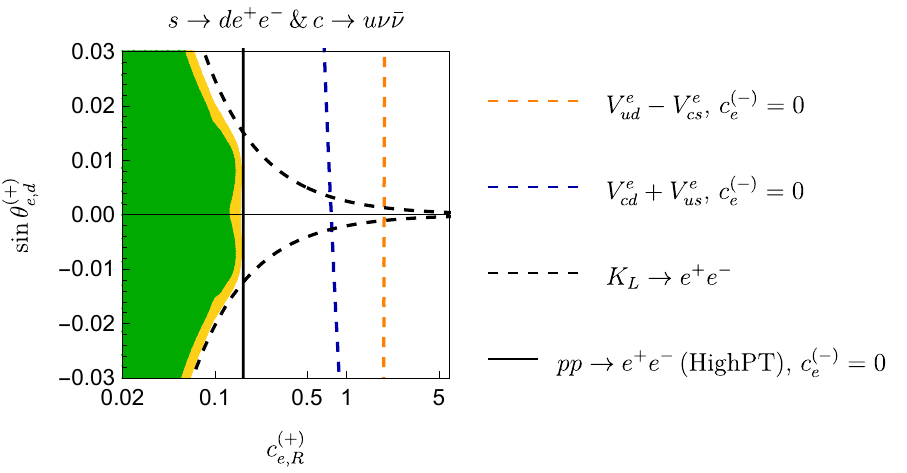}\\
    \includegraphics[scale=1.29]{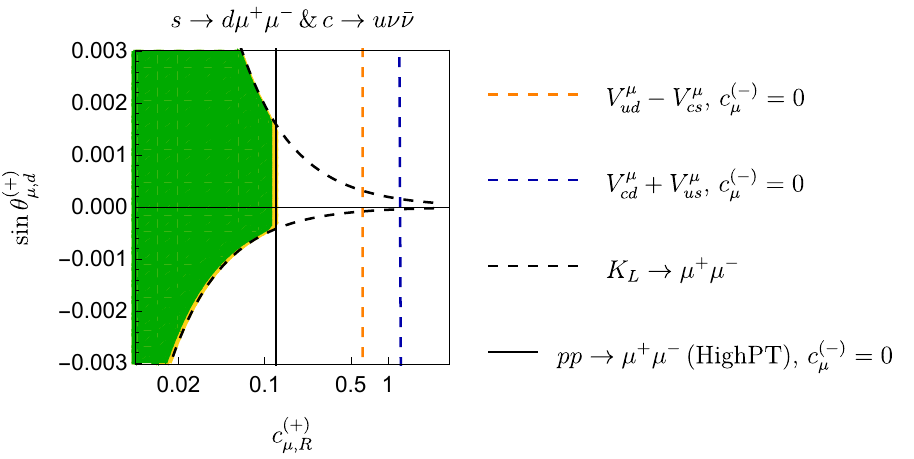}\\
    \includegraphics[scale=0.85]{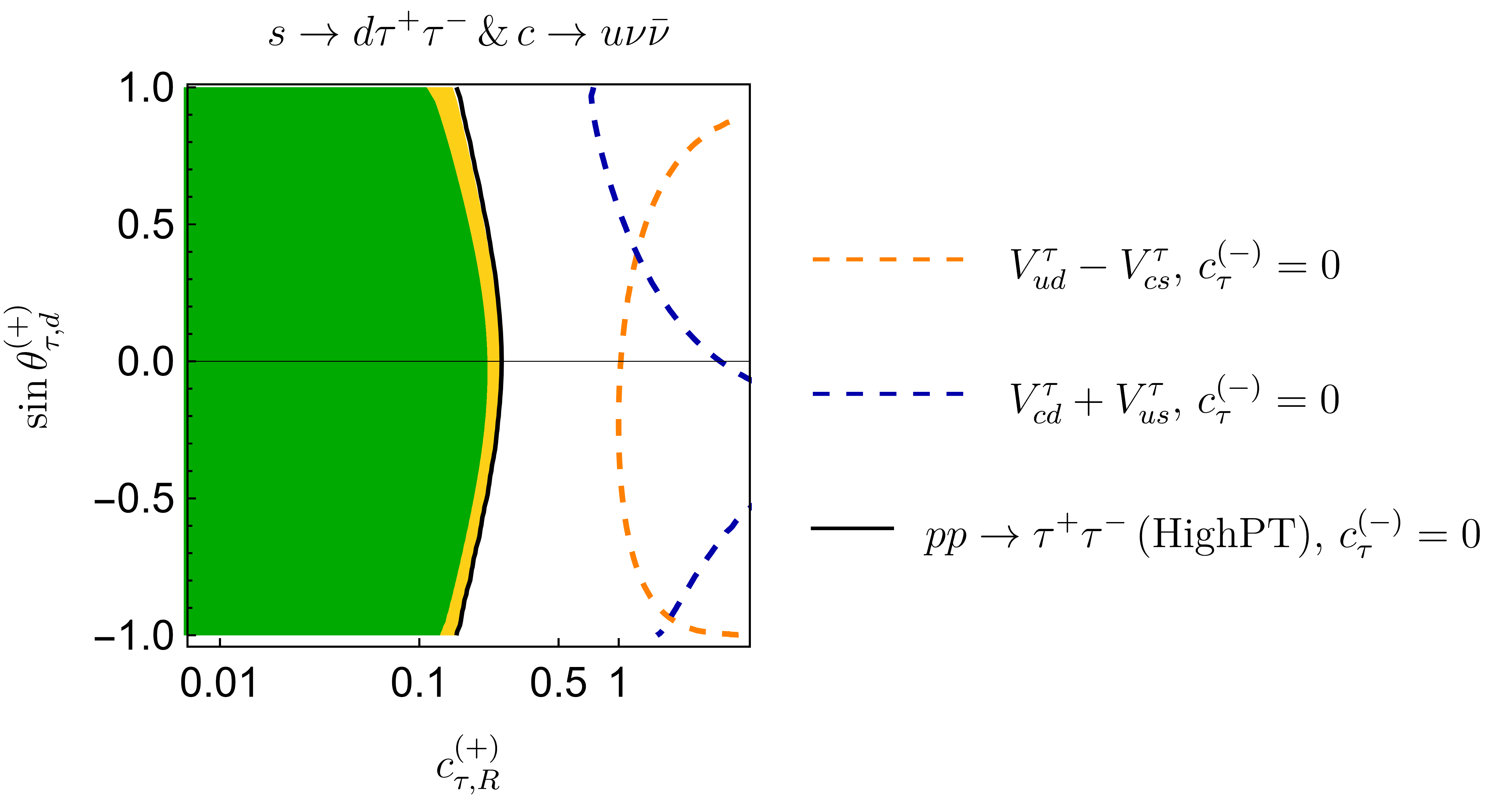}
  \end{tabular}
  \caption{Constraints on $c_R^{(+)}$ and $\theta_d^{(+)}$ from
    various experimental bounds. Here we have set $c_{\ell,I}^{(+)}$
    to zero. The green and yellow shaded regions correspond to $68\%$ CL and $95\%$ CL allowed regions of the global fit. The HighPT and charged current bounds were derived under the assumption that $c^{(-)}_{\ell, R} = c^{(-)}_{\ell, I} = 0$. }
  \label{fig:cPlusLimits2}
\end{figure}

Our main results in terms of current experimental constraints on the $X^{(\pm)}$ components are summarized in Eqs.~\eqref{eq:cIminus} and~\eqref{eq:CIplus} for the flavor universal CPV contributions, as well as in Figs.~\ref{fig:cRtanThetaMinus} and~\ref{fig:cPlusLimits2} for the CP conserving effects.

In Fig.~\ref{fig:cRtanThetaMinus} we present the combined fit to the most relevant experimental constraints on the $(-)$ operators with electrons (upper plot), muons (middle plot) and taus (lower plot). We observe that away from the down-quark mass basis alignment limit ($\theta^{(-)}_{\ell,d} \simeq 0$) the constraints are completely dominated by the NA62 measurement of $\mathcal B (K^+ \to \pi^+ \nu \bar \nu)$ (marked with black dashed lines). Thus future planned improvements in this measurement~\cite{Goudzovski:2022vbt} are expected to have an important effect on all three lepton-specific operators. 
The nontrivial behavior of the (green shaded) $68\%$ CL regions of the global fit is also due to the possible interferences between the SM and NP contributions to this decay.   
Interestingly, and as first pointed out in Ref.~\cite{Fuentes-Martin:2020lea}, the constraints in the charm sector are currently dominated by Drell-Yan measurements at the LHC (marked with full black lines), with the exception of muonic operators, where the current best constraint is given by the LHCb upper bound on $\mathcal B(D^0 \to \mu^+ \mu^-)$~\cite{LHCb:2022uzt} (marked in black dotted line). In light of this, future improvements in the search for this rare decay by both LHCb and BelleII~\cite{Belle-II:2018jsg} are thus highly anticipated (projections shown in blue dotted line). For electron operators current bounds from high-$p_T$ and rare $D\to \pi e^+ e^- $ measurements are comparable. Future measurements of the later decays by LHCb and BelleII, especially away from the long-distance resonance peaks in the $e^+ e^-$ invariant mass spectrum, could potentially improve this bound considerably. Finally, since all low energy decay channels for tauonic operators are closed, any future improvements in this sector will necessarily rely on precise  (HL)LHC measurements of the $pp\to \tau \tau$ spectrum. Currently, the high-$p_T$ experiments allow us to set a limit on the CP violating phase for the tau. The weakest derived bound reads: 
\begin{equation}
	\big|\im [c_{\tau,I}^{(+)}]\big| \lesssim 0.15\,.
\end{equation}

In Fig.~\ref{fig:cPlusLimits2} we present the combined fit to the most relevant experimental constraints on the $(+)$ operators with electrons (upper plot), muons (middle plot) and taus (lower plot). 
In this case we observe that away from the down-quark mass basis alignment limit the constraints in the electron and muon sectors are dominated by $K_L \to \ell^+ \ell^-$ decay rate measurements (marked with dashed black lines). 
Therefore it is important that in the future, a combined analysis of $K^0 \to \ell \ell$ decays~\cite{Dery:2021mct} could possibly go beyond the current sensitivity. Again high-$p_T$ Drell-Yan production measurements (marked with black full lines) are most restrictive close to $\theta^{(+)}_{\ell,d} \simeq 0$. In the case of tauonic operators, LHC constraints dominate over the whole $\theta^{(+)}_{\tau,d}$ range. Interestingly, the almost flat behavior of these constraints with $\theta^{(+)}_{\ell,d}$ is a result of the non-trivial interplay between flavor changing (${\bar s d}$ and $\bar d s$) and flavor conserving ($\bar d d$ and $\bar s s$) initial state contributions which exhibit opposite behavior, combined with the marginalization over the trace contributions ($\lambda^{(+)}$), see Eq.~\eqref{eq:highPToperators}. \\
To be competitive with the high-$p_T$ constraints, the current experimental bound on $\mathcal B (D^0 \to \pi^0 \nu\bar\nu)$ (not shown, see Sec.~\ref{sec:dpinunu}) would have to be improved by 3 orders of magnitude. 
It is also important to note that at present, neutral current constraints are already stringent enough to make possible effects in charged current transitions negligible.

%
\section{Conclusions}
\label{sec:conclusions}
%
We considered the effects of BSM physics in rare semileptonic $\Delta C=1$ and $\Delta S=1$ processes mediated by purely left-handed quark and lepton operators. 
Restricting the discussion to the two light quark generations allowed us to parametrize possible BSM effects in quark flavor space in terms of Hermitian matrices $(X)$ of
dimension two, parametrized by three real and one imaginary coefficient. 
In addition, weak isospin singlet and triplet operator contributions are split into two distinct phenomenological sectors:
one is characterized by effective BSM couplings $X_{}^{(-)}$ and contributes to
$s \to d \nu \bar \nu$  and $c \to u \ell^{+} \ell^{-}$ transitions. On the other hand $c \to u \nu \bar \nu$  and $s \to d \ell^{+} \ell^{-}$  processes can receive contributions from effective $X_{}^{(+)}$ couplings. 
A distinct feature that emerges in such a framework is that beyond the SM there exists a single universal source of CP violation, parametrized by as single CP-violating coefficient for
each ($(-)$ and $(+)$) sector.  

To determine the allowed parameter space, in the $(-)$ sector we
considered exclusive decays $D^{} \to \pi^{} \ell^{+} \ell^{-}$,
$D\to \ell^{+} \ell^{-}$ and $K^{} \to \pi^{} \nu \bar \nu$. 
The $K_{L }\to \pi^{-} \nu \bar \nu$ decay amplitude is CP violating and the existing upper bound on the corresponding decay rate from KOTO directly constrains CPV in this sector.
In practice however, the decay rate $K^{+}\to \pi^{+} \nu \bar \nu$ is currently more sensitive and already constrains the CPV
contribution in $X^{(-)}$ to below $\sim 10^{-4}$. The same
CP-violating coefficient contributes also to charm meson
decays. However, the lack of
high-precision measurements in rare charm semileptonic decays currently precludes any competitive CPV probes in this sector. 
In the $X^{(+)}$ sector, the decay modes $K_{L} \to \pi^{0 }\ell^{+} \ell^{-}$ and
$K_{L} \to\ell^{+} \ell^{-}$ dominate the low energy constraints. 

Importantly, low energy data from exclusive $K$ and $D$ decays is
complemented by constraints from high-p$_{T}$ processes
$pp\to \ell^{+} \ell^{-}$ at the LHC. In the case of electrons, these
bounds are currently competitive with the existing data on the rates of
$D^{+}\to \pi^{+} e^{+}e^{-}$ (for $X^{(-)}$) and $K_L \to e^+ e^-$ (for $X^{(+)}$). They also dominate the constraints in the $X^{(+)}$ sector for tau leptons.
Interestingly, flavor conserving and flavor changing neutral currents contribute to $pp\to \ell^{+} \ell^{-}$ in a complementary way, allowing to completely constrain both $X^{(\pm)}$ parameter spaces using only high-$p_T$ data.

The presence of weak isospin triplet operators in the effective Lagrangian
implies that charged current processes might receive BSM contributions
as well. Therefore, our analysis considered
constraints from super-allowed beta decays, from (semi)leptonic kaon decays used in the extraction of the CKM matrix element
$V_{us}$, as well as from charged current-induced (semi)leptonic decays of charmed mesons. We
found that these constraints are more pronounced in the $X^{(+)}$
sector, where they currently supersede the ones coming from the FCNC process $D \to \pi \nu \nu$. They are however not competitive with high-$p_T$ constraints.

In the future, improved bounds on BSM physics entering $D$ and $K$ rare semileptonic decays are expected from high precision measurements of both charm decay rates ($D^{+} \to \pi^{+} \ell^{+} \ell^{-}$, $D\to \ell^{+} \ell^{-}$ and $D^{} \to \pi^{} \nu \bar \nu$), as well as kaon decay rates ($K^{+} \to \pi^{+} \nu \bar \nu$,  $K_{L }\to \pi^{0} \nu \bar \nu$ and $K^0 \to \mu^+ \mu^-$). 
At the same time, future precision measurements  of high p$_{T}$ processes $pp \to \ell^{+} \ell^{-}$ at (HL)LHC, especially in the tau sector, could further illuminate and constrain possible BSM physics in strange and charm semileptonic processes.

%
\section*{Acknowledgements}
%

J.F.K. is grateful to the Mainz Institute for Theoretical Physics (MITP) of the Cluster of Excellence PRISMA${}^{+}$ (Project ID 39083149), for its hospitality and its partial support during the completion of this work. We acknowledge the financial support from the Slovenian Research Agency (research core funding No. P1-0035 and J1-3013).

\appendix

\section{Lepton-specific CKM elements}
\label{app:CKMextraction}
The lepton-specific modifications of the CKM elements of Eq.~\eqref{eq:chargedcurrentL} are:
\begin{subequations} 
\label{eq:CKMmods}
  \begin{align}
(\Lambda^2/v^2)  \Delta V^{(\ell)}_{ud} &= +\lambda^{(3,\ell)} \cos \theta_c - \frac{1}{2} \cx{+}_R \cos\left(\theta_c + \theta_d^{(+)}\right) + \frac{1}{2} \cx{-}_R \cos\left(\theta_c + \theta_d^{(-)}\right) + \frac{i}{2}\sin \theta_c \,\left(\cx{+}_I - \cx{-}_I\right) \,,\\
(\Lambda^2/v^2)  \Delta V^{(\ell)}_{us} &= +\lambda^{(3,\ell)} \sin \theta_c + \frac{1}{2} \cx{+}_R \sin \left(\theta_c + \theta_d^{(+)}\right) - \frac{1}{2} \cx{-}_R \sin\left(\theta_c + \theta_d^{(-)}\right) - \frac{i}{2}\cos \theta_c \,\left(\cx{+}_I - \cx{-}_I\right) \,,\\
(\Lambda^2/v^2)  \Delta V^{(\ell)}_{cd} &= -\lambda^{(3,\ell)} \sin \theta_c + \frac{1}{2} \cx{+}_R \sin \left(\theta_c + \theta_d^{(+)}\right) - \frac{1}{2} \cx{-}_R \sin\left(\theta_c + \theta_d^{(-)}\right) + \frac{i}{2}\cos \theta_c \,\left(\cx{+}_I - \cx{-}_I\right) \,,\\
(\Lambda^2/v^2)  \Delta V^{(\ell)}_{cs} &= +\lambda^{(3,\ell)} \cos \theta_c + \frac{1}{2} \cx{+}_R \cos \left(\theta_c + \theta_d^{(+)}\right) - \frac{1}{2} \cx{-}_R \cos\left(\theta_c + \theta_d^{(-)}\right) - \frac{i}{2}\sin \theta_c \,\left(\cx{+}_I - \cx{-}_I\right)\,,
\end{align}
\end{subequations}
As we do not assume lepton flavor universality, it is necessary to
extract constraints separately for each lepton flavour.
\subsection{$V_{ud}^{\ell}$}
For electrons, we use the measured value of the superallowed $\beta$ decay, which limits the parameter space in the direction of $C_{ud}^{e}$.
Using the results presented in \cite{1411.5987, 2002.07184}, and the derived matching conditions, we have: 
\begin{equation}
	\left| V_{ud}^e \right|^2 = \left|V_{ud}  +\frac{v^2}{\Lambda^2} \left( \lambda^{(3,e)} \cos \theta_c - \frac{1}{2} \cx{+}_R \cos\left(\theta_c + \theta_d^{(+)}\right) + \frac{1}{2} \cx{-}_R \cos\left(\theta_c + \theta_d^{(-)}\right) \right) \right|^2 = \frac{2984.432(3) s}{\mathcal{F}t(1 + \Delta_R^{V})}
\end{equation}
where  $\mathcal{F}t = 3072(2)s$ and $\Delta_R^V = 0.02426(32)$ \cite{1907.06737}\footnote{Other measurements for these values exist as well, see e.g. \cite{1807.10197}. We choose the ones with larger uncertainties.}.\\
In order to limit this CKM element for muons and taus, we use the following LFU ratio \cite{Pich:2013lsa}:
\begin{equation}\label{eq:RpiTauMu}
	R^{\pi}_{\tau/\mu} = \frac{\Gamma(\tau^- \to \pi^- \nu_\tau)}{\Gamma(\pi^- \to \mu^- \nu_\mu)} = \Bigg(\frac{V_{ud}^\tau}{V_{ud}^\mu} \Bigg)^2 \frac{m_\tau^3}{2 m_\pi m_\mu^2} \frac{\big(1 - m_\pi^2/m_\tau^2\big)^2}{\big(1 - m_\mu^2/m_\pi^2\big)}\big(1 + \delta R_{\tau/\mu}^\pi\big) \,,
\end{equation}
where $\delta R^\pi_{\tau/\mu} = (0.16 \pm 0.14) \%$ is the ratio of radiative corrections \cite{Marciano:1993sh}. When deriving limits, we consider only one lepton-specific contribution at a time, while keeping the other CKM element fixed at the SM value. \\

The theoretical prediction can be compared to the ratio of measured decays:
\begin{equation}
	R^{\pi (\mathrm{exp})}_{\tau/\mu} = 0.9962 \pm 0.0027 \,.
\end{equation}
\subsection{$V_{us}^\ell$}
We employ the precisely determined LFU ratios in kaon decays, namely: 
\begin{align}
	R^{K}_{e/\mu} &= \frac{\Gamma(K^- \to e^- \bar{\nu})}{\Gamma(K^- \to \mu^- \bar{\nu})} \label{eq:ReMu} \,,\\
	R^{K}_{\tau/\mu} &= \frac{\Gamma(\tau^- \to K^- \nu)}{\Gamma(K^- \to \mu^- \bar{\nu})}  \label{eq:RtauMu} \,,
\end{align}
and adapt the same approach as previously. The ratio in Eq.~\eqref{eq:ReMu} is used in our fits to limit $V^{e}_{us}$ and $V^{\mu}_{us}$, whereas Eq.~\eqref{eq:RtauMu} is used to limit $V^{\tau}_{us}$. 
In case NP is present only in electrons, the experimental value in Eq.~\eqref{eq:ReMu} can be compared to our theoretical prediction: 
\begin{equation}
\frac{R^{K \, (\mathrm{exp})}_{e/\mu}}{R^{K \, (\mathrm{SM})}_{e/\mu}} - 1 \approx \frac{v^2}{\Lambda^2} \left[ c_R^{(+)} \sin \left(\theta_d^{(+)} + \theta_c \right) - c_R^{(-)} \sin \left(\theta_d^{(-)} + \theta_c \right) + 2 \lambda^{(3,e)} \sin \theta_c \right] \Big/ \sin\theta_c,
\end{equation}
with similar expressions for the other two lepton generations. Here: 
\begin{equation}
R^{K (\rm exp)}_{e/\mu} = (2.488 \pm 0.010) \times 10^{-5} \,, \qquad R^{K (\rm SM)}_{e/\mu} = (2.477 \pm 0.001) \times 10^{-5}.
\end{equation}
\subsection{$V_{cd}^\ell$ and $V_{cs}^\ell$}
HFLAV \cite{2206.07501} gives world-average lepton-specific
measurements (bounds) on $V_{cd}^{\mu,\tau}$ and $V_{cs}^{\mu, \tau}$
from the branching ratios $D^+ \to \ell^+ \nu$ and $D^+_s \to \ell^+
\nu$. As the bounds on electron-specific coefficients $V_{cd}^{e}, V_{cs}^{e}$ are very weak, we use other charm decays instead, in particular $D \to \pi^- e^+ \nu$ and $D \to K^- e^+ \nu$, for which the NP-dependent branching ratios read:
\begin{equation}\label{eq:chargedcharmBR}
	\frac{d \mathcal{B}}{d q^2} (D^0 \to P e^+ \nu) = \frac{G_F^2 |V^{e}_{cq^\prime}|^2}{384 m_D^3 \pi^3 \Gamma_{D}} \bigg(1 - \frac{m_e^2}{q^2} \bigg)^2 \sqrt{\lambda} \Bigg(f_{+_{D \to P}}^2(q^2) \bigg(2 + \frac{m_e^2}{q^2} \bigg) \lambda + 3 f_{0_{D \to P}}^2(q^2) \frac{m_e^2}{q^2}\big(m_{D}^2 - m_{P}^2 \big)^2 \Bigg).
\end{equation}
Here $\lambda \equiv \lambda(m_D^2, m_P^2, q^2)$ and $P = \pi^-$ ($q^\prime = d$) or $K^-$ ($q^\prime = s$). The momentum transfer squared can take values in the range $m_e^2 \leq q^2 \leq (m_D - m_P)^2$. We can compare the integrated branching ratios with the experimental values: 
\begin{equation}
\mathcal{B}^{(\rm exp)} (D^0 \to \pi^- e^+ \nu) = (2.91 \pm 0.04 ) \times 10^{-3}, \qquad \mathcal{B}^{(\rm exp)} (D^0 \to K^- e^+ \nu) = (3.549 \pm 0.026) \times 10^{-2}.
\end{equation}

\clearpage
\subsection{Limits}
We summarize our results in the Table \ref{tab:LeptonSpecificCKMtotal}. 
\vspace{1cm}
\begin{table}[!h]
	\centering
	\begin{tabular}{|c|c|c|c|}
		\hline
		& CKM Element  & Measurement/Experiment & Derived value  \\ \cline{2-4}
		& $V^{e}_{ud}$ & Superallowed $\beta$ decay &  $0.9736 \pm 0.0002$ \\
$e$		& $V^{e}_{us}$ & $R^K_{e/\mu}$ (NP in $e$)  & $0.2255 \pm 0.0004$ \\
		& $V^{e}_{cd}$ & $D^0 \to \pi^- e^+ \nu $ &  $-0.235 \pm 0.008$ \\
		& $V^{e}_{cs}$ & $D^0 \to K^- e^+ \nu $ & $0.98 \pm 0.04$ \\ \cline{2-4}
		& $V^{\mu}_{ud}$ & $R^\pi_{\tau/\mu}$ (NP in $\mu$)  & $0.9781 \pm 0.0027$ \\
$\mu$	& $V^{\mu}_{us}$ & $R^K_{e/\mu}$ (NP in $\mu$) & $ 0.2245 \pm 0.0005$  \\
		& $V^\mu_{cd}$ & $D^+ \to \mu^+ \nu$ (HFLAV) & $ -0.225 \pm 0.007$ \\
		& $V^\mu_{cs}$ & $D^+_s \to \mu^+ \nu$ (HFLAV) & $0.97 \pm 0.02$ \\ \cline{2-4}
		& $V^{\tau}_{ud}$ & $R^\pi_{\tau/\mu}$ (NP in $\tau$) & $0.9707 \pm 0.0026$ \\
$\tau$	& $V^{\tau}_{us}$ & $R^K_{\tau/\mu}$ (NP in $\tau$) & $0.222 \pm 0.002$ \\
		& $V^\tau_{cd}$ & $D^+ \to \tau^+ \nu$ (HFLAV) & $-0.25 \pm 0.03$ \\
		& $V^\tau_{cs}$ & $D^+_s \to \tau^+ \nu$ (HFLAV) & $0.98 \pm 0.02$\\
		\hline
	\end{tabular}
\caption{Experimental input used to limit the lepton-specific CKM contributions.}
\label{tab:LeptonSpecificCKMtotal}
\end{table}

\bibliographystyle{apsrev}
\bibliography{draft}
\end{document}